\documentclass[aps,prb,twocolumn,letterpaper,longbibliography,superscriptaddress,floatfix,amssymb,amsmath]{revtex4-1}

\usepackage{graphicx}
\usepackage{amsmath,amssymb}
\usepackage[resetlabels]{multibib}
    \newcites{S}{supp}
\usepackage[colorlinks,urlcolor=black,citecolor=black,linkcolor=black]{hyperref}

\makeatletter
\@addtoreset{figure}{myfigure}
\def\@caption@fignum@sep{~$\vert$~}%
\def\fnum@figure{\textbf{\figurename~\thefigure}}
\makeatother

\newcommand*{\citen}[1]{%
  \begingroup
    \romannumeral-`\x 
    \setcitestyle{numbers}%
    \cite{#1}%
  \endgroup
}

\renewcommand{\figurename}{Figure}

\newcommand{\micron}{\ensuremath{\mu\mathrm{m}}}

\newcommand{\threevec}[3]{\left(\begin{array}{c}#1\\#2\\#3\end{array}\right)}

\newcommand{\bvec}[1]{\mathbf{#1}}

\newcommand{\ket}[1]{\ensuremath{\left|#1\right>}}

\newcommand{\inleva}[1]{\langle#1\rangle}
\newcommand{\abs}[1]{\left|#1\right|}

\newcommand{\SO}{\ensuremath{\mathrm{SO}}}

\newcommand{\U}{\ensuremath{\mathrm{U}}}

\newcommand{\nematic}{\mathbf{\hat{d}}}
\newcommand{\absF}{\ensuremath{|\langle\bvec{\hat{F}}\rangle|}}
\newcommand{\expF}{\ensuremath{\langle\bvec{\hat{F}}\rangle}}

\newcommand{\rhohat}{\ensuremath{\boldsymbol{\hat{\rho}}}}

\newcommand{\xhat}{\ensuremath{\mathbf{\hat{x}}}}
\newcommand{\yhat}{\ensuremath{\mathbf{\hat{y}}}}
\newcommand{\zhat}{\ensuremath{\mathbf{\hat{z}}}}
\newcommand{\nhat}{\ensuremath{\mathbf{\hat{n}}}}
\newcommand{\rr}{\ensuremath{\mathbf{r}}}
\newcommand{\nuke}[1]{}
\newcommand{\polar}{P}

\newcommand{\beq}{\begin{equation}}
\newcommand{\eeq}{\end{equation}}

\begin{document}

\author{L.~S.~Weiss} 
\altaffiliation{Johns Hopkins University Applied Physics Laboratory, 11100 Johns Hopkins Road, Laurel, MD 20723, USA}
\affiliation{Department of Physics and Astronomy, Amherst College, Amherst, Massachusetts 01002--5000, USA}

\author{M.~O.~Borgh} 
\affiliation{Faculty of Science, University of East Anglia, Norwich, NR4 7TJ, United Kingdom}

\author{A.~Blinova} 
\affiliation{Department of Physics, University of Massachusetts, Amherst, Massachusetts 01003, USA}
\affiliation{Department of Physics and Astronomy, Amherst College, Amherst, Massachusetts 01002--5000, USA}

\author{T.~Ollikainen} 
\affiliation{QCD Labs, QTF Centre of Excellence, Department of Applied Physics, Aalto University, P.O. Box 13500, FI--00076 Aalto, Finland}
\affiliation{Department of Physics and Astronomy, Amherst College, Amherst, Massachusetts 01002--5000, USA}

\author{M.~M\"ott\"onen} 
\affiliation{QCD Labs, QTF Centre of Excellence, Department of Applied Physics, Aalto University, P.O. Box 13500, FI--00076 Aalto, Finland}
\affiliation{VTT Technical Research Centre of Finland Ltd, P.O. Box 1000, FI-02044 VTT, Finland}

\author{J.~Ruostekoski} 
\affiliation{Department of Physics, Lancaster University, Lancaster, LA1 4YB, United Kingdom}

\author{D.~S.~Hall} \email{dshall@amherst.edu (corresponding author)}
\affiliation{Department of Physics and Astronomy, Amherst College, Amherst, Massachusetts 01002--5000, USA}

\title{Controlled Creation of a Singular Spinor Vortex \\ by Circumventing the Dirac Belt Trick}

\date{September 25, 2019}


\begin{abstract}
Persistent topological defects and textures are particularly dramatic consequences of superfluidity. Among the most fascinating examples are the singular vortices arising from the rotational symmetry group $\SO(3)$, with surprising topological properties illustrated by Dirac's famous belt trick. Despite considerable interest, controlled preparation and detailed study of vortex lines with complex internal structure in fully three-dimensional spinor systems remains an outstanding experimental challenge. Here, we propose and implement a reproducible and controllable method for creating and detecting a singular $\SO(3)$ line vortex from the decay of a non-singular spin texture in a ferromagnetic spin-1 Bose--Einstein condensate. Our experiment explicitly demonstrates the $\SO(3)$ character and the unique spinor properties of the defect. Although the vortex is singular, its core fills with atoms in the topologically distinct polar magnetic phase. The resulting stable, coherent topological interface has analogues in systems ranging from condensed matter to cosmology and string theory.
\end{abstract} 

\maketitle


\begin{center}
\textbf{\large Introduction}
\end{center}

Quantized vortices are topological defects with universal properties that span seemingly disparate areas of science, such as high-energy physics, superconductivity, liquid crystals, and superfluids\cite{volovik}. Superfluids with internal degrees of freedom such as liquid $^3$He$^{\,}$ (Ref.~\citen{vollhardt-wolfle}) and dilute-gas spinor Bose--Einstein condensates\cite{kawaguchi_physrep_2012,stamper-kurn_rmp_2013} (BECs) may exist in diverse stable phases, characterised by different broken symmetries of the full interaction Hamiltonian.
Distinct states within a given phase of matter transform into one another in several ways, such as through rotations of spin and condensate phase. As a result, a rich phenomenology emerges consisting of topological defects and textures that resemble those predicted to exist in quantum field theories and cosmology\cite{volovik}.

In ordinary scalar superfluids, such as superfluid liquid $^4$He and dilute Bose--Einstein condensates with frozen internal degrees of freedom, quantized vortices are characterised by the winding of the phase of the macroscopic wave function about any closed loop encircling the vortex line\cite{donnelly,fetter_rmp_2009}. The whole spectrum of phase values converges to the singular vortex line, at which the superfluid density vanishes.

In contrast, spin-1 condensates are described by a three-component spinor, which admits both polar (\polar) and ferromagnetic (FM) ground-state magnetic phases. For atoms in the FM phase, the magnitude of the spin assumes its maximum value of one\cite{kawaguchi_physrep_2012,stamper-kurn_rmp_2013}, and all of the different physical states are related to each other by spatial rotations of the spinor. The distinguishable states of the system are fully represented by the orientation of a local, orthonormal vector triad defined by the orientation of the atomic spin and rotations about it, corresponding to the elements of the group $\SO(3)$ of three-dimensional (3D) spatial rotations. Mathematical analysis\cite{mermin_rmp_1979} of this symmetry group indicates that vorticity must be carried either by coreless, non-singular spin textures, or by quantized, singular vortices.

Quantized singular $\SO(3)$ vortices with even winding numbers have the unusual property that they are topologically equivalent to the defect-free state. When the local orientation of the vector triad describing the $\SO(3)$ vortex undergoes an even number of $2\pi$ rotations about an axis passing through the system, the triads can be locally and continuously reoriented, smoothly returning the system to a uniform configuration. Equivalently, joining two vortices with $2\pi$ winding each can cancel their net vorticity, either when they circulate oppositely or --- less obviously ---  when they wind in the same sense. This essential property has been attributed to Dirac as his eponymous belt trick, a demonstration in which two $2\pi$ twists of a belt in the same direction may return it to its original configuration\cite{staley_ejp_2010}; but the concept also makes an appearance in diverse artistic contexts such as the Balinese candle dance. The significance of the belt trick to our work is that vortices with an odd number of $2\pi$ rotations of the vector triad are all equivalent to one another but not to the defect-free state.

In light of their peculiar properties, which have no correspondence in scalar quantum fluids, singular $\SO(3)$ vortices have attracted considerable attention in several different contexts. They have previously been described and indirectly detected in the superfluid liquid $^3$He-$A$ phase~\cite{Simola87,parts_prl_1995,vollhardt-wolfle}, where their direct visualisation is challenging. In spin-1 BECs, they have been studied theoretically as the unique class of singular vortices in the FM phase\cite{isoshima_pra_2002,lovegrove_pra_2012,kobayashi_pra_2012,lovegrove_pra_2016}. Of particular significance is the fact that, although the superfluid density in the FM phase must vanish along the line where the triad orientation is ill-defined, the singular vortex can lower its energy by developing a superfluid core consisting of atoms in the spinless \polar\ phase that are excited out of the FM ground-state manifold\cite{ruostekoski_prl_2003,lovegrove_pra_2012,lovegrove_pra_2016}. This phenomenon has been observed experimentally in the spontaneous vorticity of randomly appearing singular $\SO(3)$ defects in quasi-two-dimensional (2D) condensates during a rapid non-equilibrium phase transition\cite{sadler_nature_2006}, where the filled vortex cores were detected indirectly by their lack of longitudinal magnetisation. More recently, atomic condensates subjected to momentum-dependent artificial gauge potentials were shown to support filled-core vortices\cite{chen_prl_2018} closely related to those studied in our work. Related but topologically different half-quantum vortices have also been observed in the \polar\ phase in a quasi-2D BEC\cite{seo_prl_2015}. Despite these efforts, the controlled creation of singular $\SO(3)$ vortices remains an experimental challenge.

\begin{figure}
    \centering
    \includegraphics[width=\columnwidth]{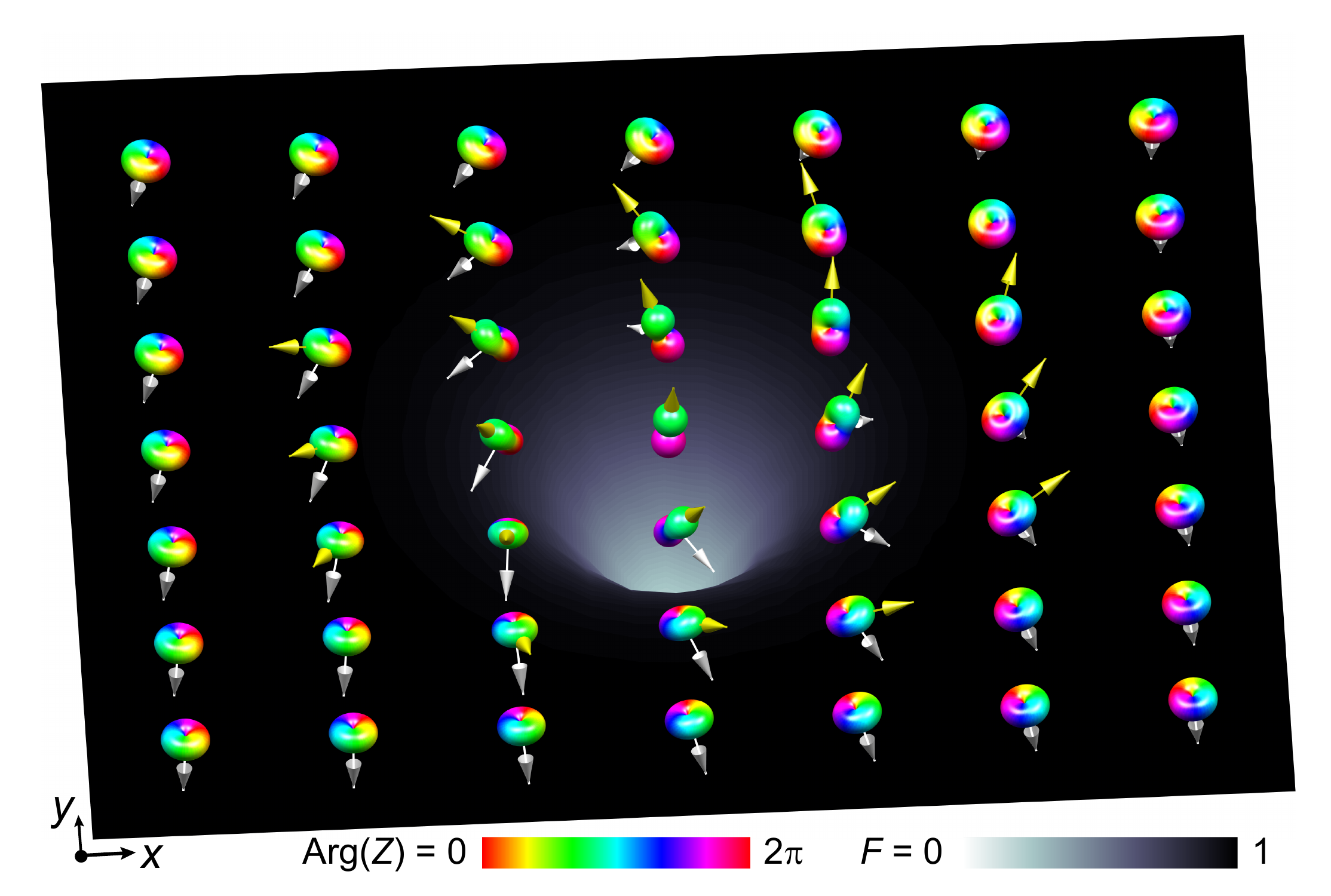}
    \caption{\textbf{Numerical simulation of a singly-quantized singular SO(3) vortex.}
    The symmetry of the local spinor order parameter is illustrated by a graphical representation of the surface of $|Z(\theta,\phi)|^2$, with the colour indicating $\mathrm{Arg}(Z)$, where $Z(\theta,\phi)=\sum_{m=-1}^{+1}Y_{1,m}(\theta,\phi)\zeta_m$ corresponds to an expansion of the spinor in terms of spherical harmonics $Y_{1,m}(\theta,\phi)$, such that $(\theta,\phi)$ define the local spinor orientation. Outside the vortex core the order parameter reaches the $\SO(3)$ symmetric ferromagnetic phase. Inside the core it continuously connects with the nematic order parameter of the polar  phase
    at the vortex line singularity, forming a coherent topological interface. The interpolation of the order parameter across the interface is readily apparent in the vanishing magnitude of the spin vector $\expF$ (silver arrows and background surface) at the vortex line
    where the polar  phase is determined by the $\nematic$ vector (gold arrows) and the phase of the macroscopic condensate wave function. Source data are provided as a Source Data file.}
    \label{fig:newintroduction}
\end{figure}

Here, in a striking manifestation of the topological constraints of the $\SO(3)$ order parameter, we transform a non-singular vortex that is topologically equivalent to one with a 4$\pi$ winding of the FM order parameter into a pair of spatially-separated singular $\SO(3)$ vortices with 2$\pi$ winding each (Fig.~\ref{fig:newintroduction}). We thereby circumvent the smooth topological unwinding permitted by Dirac's belt trick, dividing the equivalent of a $4\pi$ rotation into two $2\pi$ rotations that, once separated, cannot individually unwind. We find experimentally that the singular FM vortex cores are filled and expanded by atoms in the \polar\ phase. This establishes the existence of a coherent topological interface\cite{borgh_prl_2012,lovegrove_pra_2016}, where the order parameter continuously interpolates between the two magnetic phases within the vortex core.  Such topological interfaces are universal across many areas of physics, including superfluid liquid $^3$He at the boundary between coexisting $A$ and $B$ phases\cite{finne_rpp_2006,bradley_nphys_2008}, early-universe cosmology and superstring theory as domain walls\cite{kibble_jpa_1976} and branes\cite{sarangi_plb_2002}, and solid-state physics supporting exotic superconductivity\cite{bert_nphys_2011}.  Finally, we explicitly demonstrate the $\SO(3)$ character of the vortices by enacting a change of basis, which appears experimentally as a spatial separation of phase singularities in the three spinor components. Our work directly addresses the challenges of controlled creation and simple parameter tuning of a fully 3D, singular $\SO(3)$ vortex, marking the path for a detailed study and direct imaging of the underlying topological phenomena.


\begin{center}
\textbf{\large Results}
\end{center}

\textbf{Theoretical background.} The macroscopic wave function of a spin-1 BEC can be
written  in terms of the atomic
density $n$
and the three-component spinor $\zeta$ as $\Psi(\rr,t)=\sqrt{n(\rr,t)}\zeta(\rr,t)$.
In the FM phase,  we have \cite{kawaguchi_physrep_2012}
\begin{equation}
  \label{eq:general-fm}
  \zeta^{\mathrm{FM}} = \frac{e^{-i\gamma}}{\sqrt{2}}
    \threevec{\sqrt{2}e^{-i\alpha}\cos^2\frac{\beta}{2}}
             {\sin\beta}
             {\sqrt{2}e^{i\alpha}\sin^2\frac{\beta}{2}},
\end{equation}
which can be obtained by applying a 3D spin rotation $U(\alpha,\beta,\gamma)$ to the representative FM spinor $(1,0,0)^\mathrm{T}$. Any FM spinor is thus fully specified
by the three Euler angles $\alpha$, $\beta$, and $\gamma$, corresponding to the group of rotations in three dimensions, $\SO(3)$.
As a consequence, any FM state can be represented by the orientation of a vector triad defined by the condensate spin vector $\expF$ ($F\equiv\absF=1$) and an orthogonal vector
$\nematic $ (Methods).

\begin{figure*}
    \centering
    \includegraphics[width=0.75\linewidth]{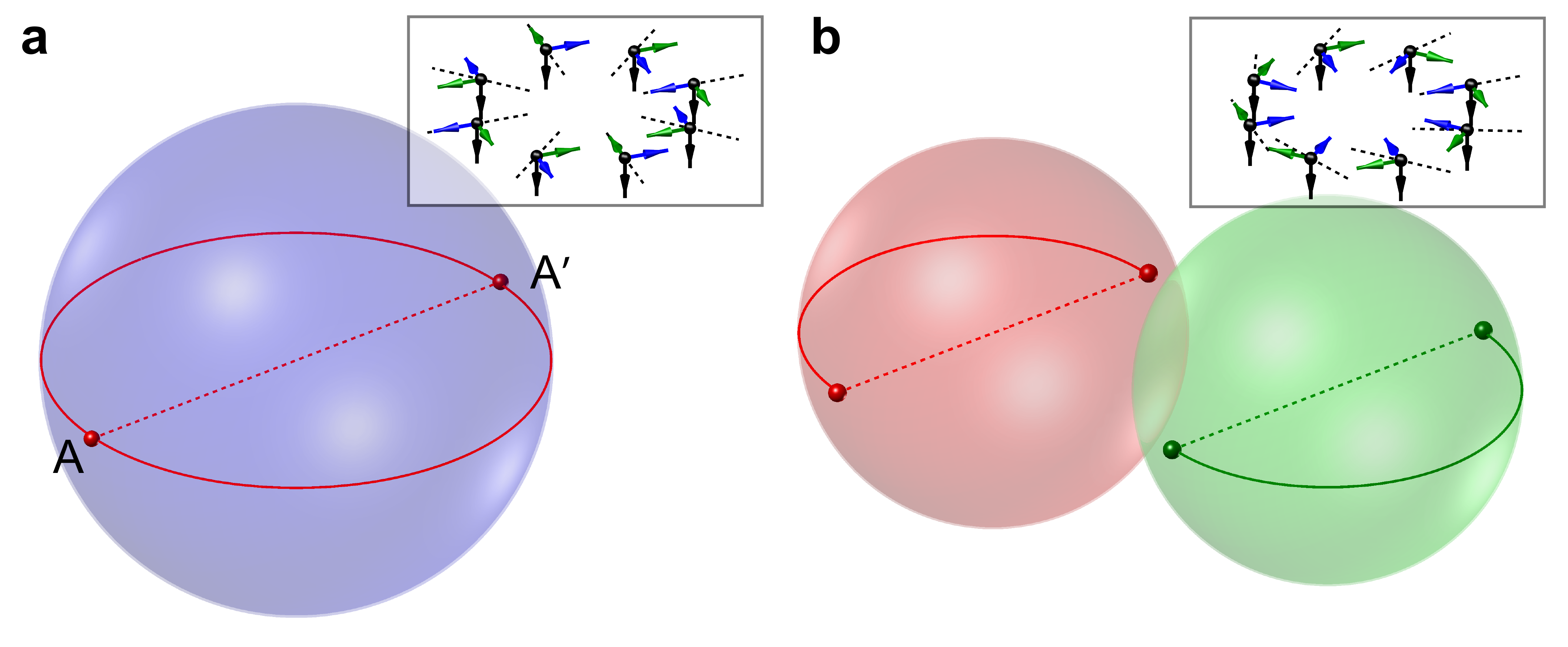}
    \caption{\textbf{Contractible and non-contractible loops in SO(3).}
    (\textbf{a}) Points inside and on the surface of the sphere represent elements of $\SO(3)$, with diametrically opposite points (e.g., $A$ and $A'$) on the surface corresponding to the same element. The contractible loop on the surface of the sphere corresponds to a vortex with $4\pi$ winding that is continuously deformable into the vortex-free state. Such deformation amounts to enacting the Dirac belt trick.  (\textbf{b}) The decay of the non-singular vortex into two separated singular line defects is represented by the emergence of two loops in distinct copies of the $\SO(3)$ sphere. Each loop is closed by virtue of the identification of $A$ and $A'$ and cannot be contracted to a point on its own.  The insets show the orientation of the ferromagnetic order parameter in real space corresponding to the points on the contractible (\textbf{a})  and non-contractible (\textbf{b}) loops, respectively. Each orthonormal triad is specified in terms of its spin direction (black arrows) and two other mutually orthonormal vectors (green and blue), with an axis of rotation given by a dashed line.}
    \label{fig:solid-sphere}
\end{figure*}

The topological stability of a singular $\SO(3)$ vortex is characterised by the way closed contours encircling the defect map into the order parameter space\cite{mermin_rmp_1979}. If the order parameter space image of such a closed loop can be continuously contracted to a point, the defect is not topologically stable against transformations to the vortex-free state. The $\SO(3)$ parameter space may be represented geometrically as a solid sphere of radius $\pi$, where the direction of the radius vector of any point within the sphere gives an axis of rotation and its length gives the rotation angle (Fig.~\ref{fig:solid-sphere}). However, $\pi$ rotations about axes $\nhat$ and $-\nhat$ are equivalent, and thus diametrically opposite points on the surface must be identified. Therefore, only two topologically distinct classes of singular vortex lines exist: those that trace between identified, diametrically opposite points an even number of times, including zero; and those that trace between them an odd number of times. Mathematically, the vortex charges form the two-element group $\mathbb{Z}_2$.


Since an even number of connections between identified points always corresponds to a loop contractible to a point, the vortices in the first (even) class can be continuously deformed into the defect-free state, and those in the second (odd) class can be continuously deformed to a singly quantized singular vortex. The essence of Dirac's belt trick is that a 4$\pi$ winding, with a path in parameter space that goes about the sphere once, is equivalent to the defect-free state.

\begin{figure*}
\centering
\includegraphics[width=\linewidth]{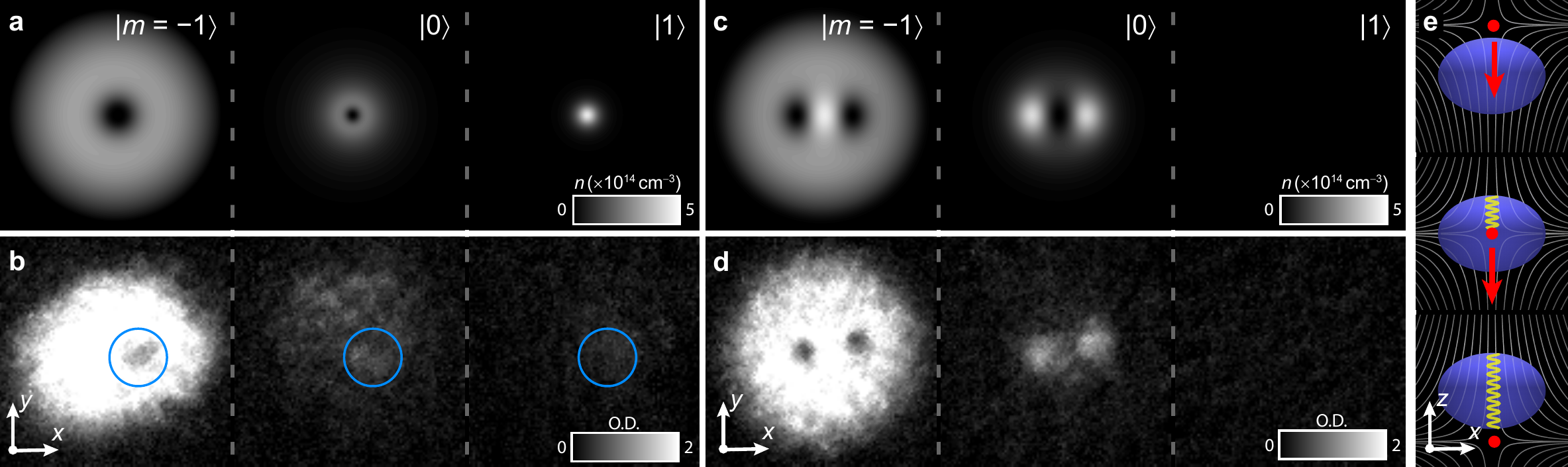}
\caption{\textbf{Controlled singular SO(3) vortex creation from a non-singular vortex.}
(\textbf{a}) Spinor component densities of an analytically constructed non-singular vortex state immediately after imprinting (see also Supplementary Note~1) in a cross-section through the condensate.
(\textbf{b}) Corresponding experimental absorption images with $T_\mathrm{evolve} \approx 0$~ms for $\mathrm{d}B_\mathrm{b}/\mathrm{d}t = -5~\mathrm{G~s}^{-1}$. The density minimum of the $m=-1$ component marks the non-singular vortex centre, and the other two components are non-zero in this region (blue circles). The creation process subjects the three spinor components to sustained differential forces, distorting the condensate and inducing non-zero densities in the $m=0$ and $m=1$ components distant from the vortex centre. The densities of the experimental images are expressed in terms of dimensionless optical depth (O.D.), and the field of view of each image is $219~\micron \times 219~\micron$.
(\textbf{c}) Locally stable state after numerically simulated energy relaxation of the non-singular vortex of \textbf{b}, shown as component densities in a cross-section through the cloud. As a consequence of the $\SO(3)$ order parameter symmetry and conserved magnetisation, the non-singular vortex is unstable towards splitting into a pair of singly quantized singular vortices, visible as density dips in the $m=-1$ component. Peaks in the $m=0$ component at the positions of the vortices show the formation of vortex cores filled with atoms in the polar phase.
(\textbf{d}) As \textbf{b}, but for $T_\mathrm{evolve} = 150$~ms and corresponding to \textbf{c}.
(\textbf{e}) Schematic of the imprinting process, showing the condensate (blue), magnetic field lines (gray), nodal line/axis of coreless vortex (yellow), and location of the magnetic field zero (red dot) at three sequential instants in time. The non-singular vortex is created by incompletely adiabatic spin rotations as the location of the field zero passes through the condensate in the direction of the red arrow. Source data are provided as a Source Data file.
\label{fig:theory+data-densities}}
\end{figure*}

\begin{figure}
  \centering
  \includegraphics[width=\columnwidth]{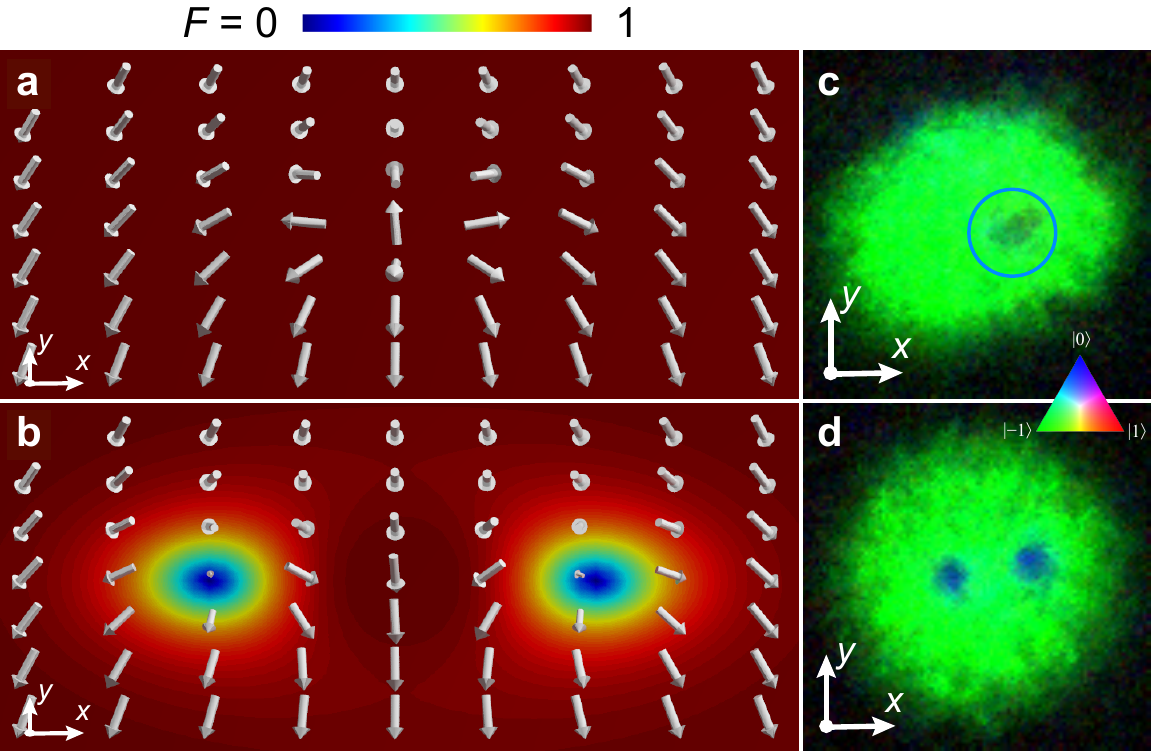}
  \caption{\textbf{Theoretical spin textures and corresponding experimental data.} (\textbf{a}) The characteristic fountain-like spin texture of the initial non-singular vortex, with spin magnitude one
    everywhere. (\textbf{b}) The spin texture of the relaxed vortex state. The background colour indicates spin magnitude, showing the filled vortex cores. (\textbf{c},\textbf{d}) Experimentally obtained composite colour images of the corresponding structures using the data of Fig.~\protect\ref{fig:theory+data-densities}, where the colours indicate the spinor components. In the absence of atoms in the $m=1$ spinor component, pure blue represents the P phase and pure green represents the ferromagnetic phase. The field of view of \textbf{c}, \textbf{d} is $219~\micron \times 219~\micron$. Source data are provided as a Source Data file.}
  \label{fig:theory+data}
\end{figure}

\textbf{SO(3) vortex creation.} Our primary result is a controlled creation method of a pair of singular $\SO(3)$ spinor vortices with non-trivial rotational topology from a non-singular texture. In the initial non-singular vortex --- also known as a coreless vortex, baby skyrmion, or Anderson--Toulouse--Chechetkin/Mermin--Ho\cite{vollhardt-wolfle} vortex in superfluid liquid helium --- the circulation is not quantized and the spin forms a fountain-like profile that adjusts to the angular momentum of the superfluid. This characteristic fountain texture has been experimentally observed in BECs\cite{leanhardt_prl_2003,leslie_prl_2009,choi_prl_2012}. If the non-singular spin texture is not constrained, e.g., energetically, it can continuously deform to a vortex-free state. We find, however, that a very sharp bending of the vortex spin profile, corresponding to a strong but incomplete longitudinal magnetisation, induces an instability wherein the non-singular spin texture decays by splitting into a pair of singly quantized vortices~\cite{lovegrove_pra_2016}, as shown in Fig.~\ref{fig:theory+data-densities}a--d (see also Supplementary Note~1). Once separated, the resulting singly quantized vortices can no longer unwind on their own, thus circumventing the Dirac belt trick along the lines of Fig.~\ref{fig:solid-sphere}. The decay paths of the non-singular vortex therefore include not only its unwinding by local spin rotations or departure from the condensate at its boundary\cite{leanhardt_prl_2003}, but also its splitting into a pair of singly quantized $\SO(3)$ vortices that will, in turn, also ultimately leave the condensate. Numerically, a bending with magnetisation $M\lesssim-0.3$ that is explicitly conserved is sufficient to guarantee the splitting, as shown in Figs.~\ref{fig:theory+data-densities} and~\ref{fig:theory+data}.

The splitting process of the non-singular spin texture is fundamentally different from the previously observed decay of a multiply quantized singular vortex into multiple singly quantized vortices\cite{leanhardt_prl_2002,shin_prl_2004,huhtamaki_prl_2006,isoshima_prl_2007}, in which magnetic trapping fields froze the atomic spin degree of freedom to produce a scalar BEC. In contrast, our experiment relies upon an all-optical trap that allows the atoms to retain their spinor nature. Even so, imprinting a multiply quantized singular vortex fully spin-polarises the condensate and spinor dynamics do not occur due to conservation of the maximised longitudinal magnetisation. The critical feature of our experiment is that the decay dynamics begin with an imprinted non-singular spin texture. The incomplete magnetisation ensures active spin degrees of freedom, and a spinor description is required. The relevant algebra of the line-vortex charges in our splitting process in $\SO(3)$ thus obeys the cyclic group $\mathbb{Z}_2$ with only the elements $0$ and $1$. Both evenly quantized and non-singular vortices are represented by the trivial element and their splitting corresponds to the group operation $0 = 1+1$, with no counterpart in a scalar BEC.

We use time-varying magnetic fields (Fig.~\ref{fig:theory+data-densities}e) to initiate the creation process experimentally with a condensate initially prepared in $\ket{m=1}$, where $|m\rangle$ denotes the $m$th spinor component. Such techniques\cite{NAKAHARA00,pietila_prl_2009_dirac} have been used to prepare, e.g., non-singular\cite{leanhardt_prl_2003,choi_prl_2012,choi_njp_2012} and multiply quantized vortices\cite{shin_prl_2004}, as well as monopoles\cite{ray_nature_2014}, skyrmions\cite{lee_sciadv_2018}, and knots\cite{hall_nphys_2016}.

Controlled creation of singular vortices in scalar BECs\cite{matthews_prl_1999,andersen_prl_2006} and continuous textures in spinor systems\cite{leslie_prl_2009} have also been achieved using phase imprinting methods.
In our experiment the atoms experience an applied magnetic field described by
\begin{equation}
  \label{eq:quadrupole-field}
  \mathbf{B} = B_\mathrm{b}(t)\zhat + b_\mathrm{q}\left(x\xhat+y\yhat-2z\zhat\right).
\end{equation}
where $b_\mathrm{q}$ is the strength of the quadrupole contribution and $B_\mathrm{b}(t)$ is a time-dependent bias field that shifts the location of the point at which the magnetic field vanishes (the field zero) to $z_0=B_\mathrm{b}/(2b_\mathrm{q})$ on the $z$-axis. We initially choose $B_\mathrm{b}$ such that the field zero is slightly above the condensate (see Methods) and the magnetic field is approximately uniform (Fig.~\ref{fig:theory+data-densities}e).

\begin{figure*}
\centering
\includegraphics[width=0.75\linewidth]{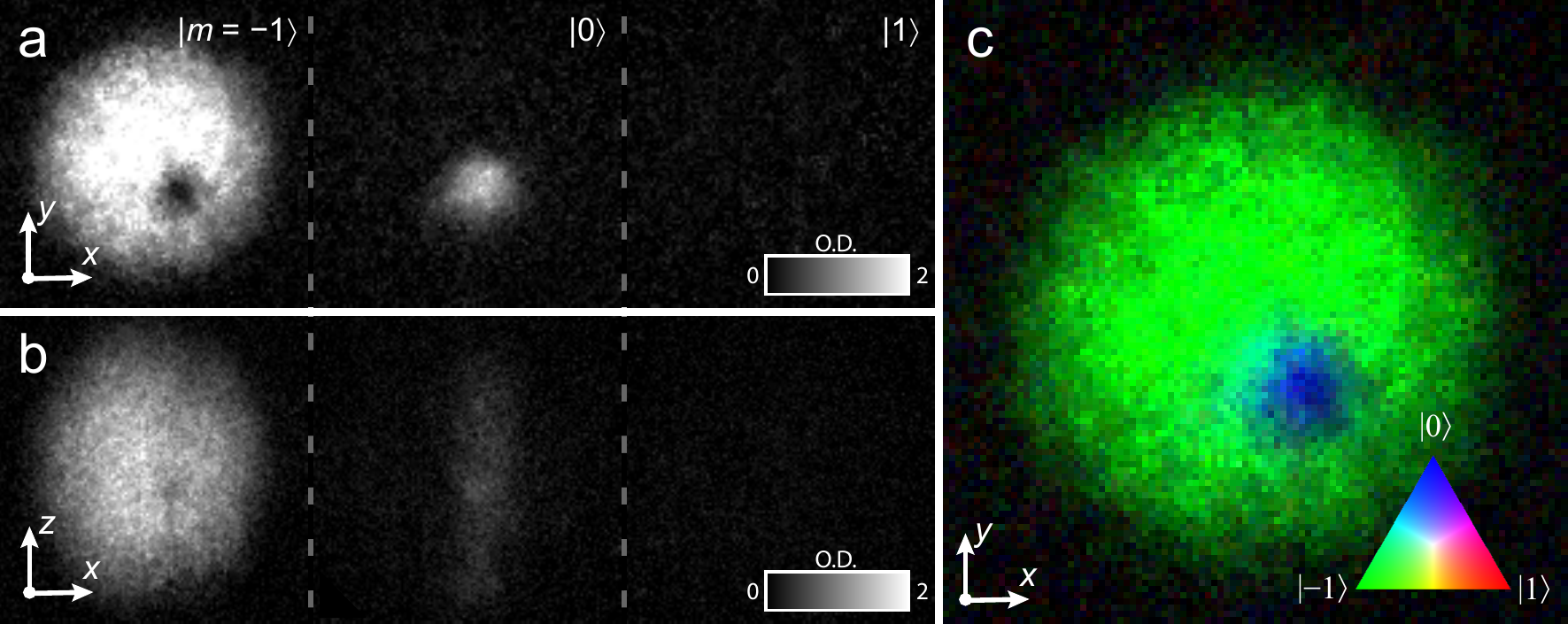}
\caption{\textbf{A singly-quantized, singular SO(3) vortex in three dimensions.}
(\textbf{a,b}) Experimental absorption images of the atomic density in each spinor component from the top, \textbf{a}, and side, \textbf{b}, expressed in terms of dimensionless optical depth (O.D.), for the $\SO(3)$ vortex configuration after one of the two vortices has exited the system. The $m=-1$ component displays a vortex line, the core of which is filled with $m=0$ atoms. The system has evolved for 1000~ms after imprinting.
(\textbf{c}) Composite false colour image of the condensate density as viewed from the top. The field of view of each image is $219~\micron \times 219~\micron$. Source data are provided as a Source Data file.}
\label{fig:mainresult}
\end{figure*}

Reducing the bias field slowly induces adiabatic spin rotations as the magnetic field zero passes through the condensate from above, trailed by a 3D nodal line\cite{ray_nature_2014} (Fig.~\ref{fig:theory+data-densities}e). At faster magnetic field ramp rates the otherwise identical experiment yields controllably incomplete adiabatic spin rotations, and results in a non-singular vortex\cite{leanhardt_prl_2002,choi_njp_2012} with additional populations in $\ket{0}$ and $\ket{1}$ (Fig.~\ref{fig:theory+data-densities}a,b and Supplementary Note~1). The atoms are released from the trap after an evolution time $T_\mathrm{evolve}$, measured from the completion of the field ramp. Following a period of ballistic expansion they are imaged, whereupon we observe a pair of singly quantized $\SO(3)$ vortices in $\ket{-1}$ with filled cores containing atoms in $\ket{0}$, as shown in Figs.~\ref{fig:theory+data-densities}c,d and~\ref{fig:theory+data}. These results agree with a numerical simulation of the locally relaxed state (Supplementary Note~1).  One of these singular spinor vortices typically departs the condensate before the other, thus lowering the condensate energy\cite{lovegrove_pra_2012,lovegrove_pra_2016} and leaving behind a single $\SO(3)$ vortex (Fig.~\ref{fig:mainresult}). The main dissipative sources, as in scalar BECs\cite{fetter_rmp_2009,rosenbusch_prl_2002}, are a non-vanishing thermal cloud and potential collisions with high-temperature atoms.



\begin{figure}
\centering
\includegraphics[width=\columnwidth]{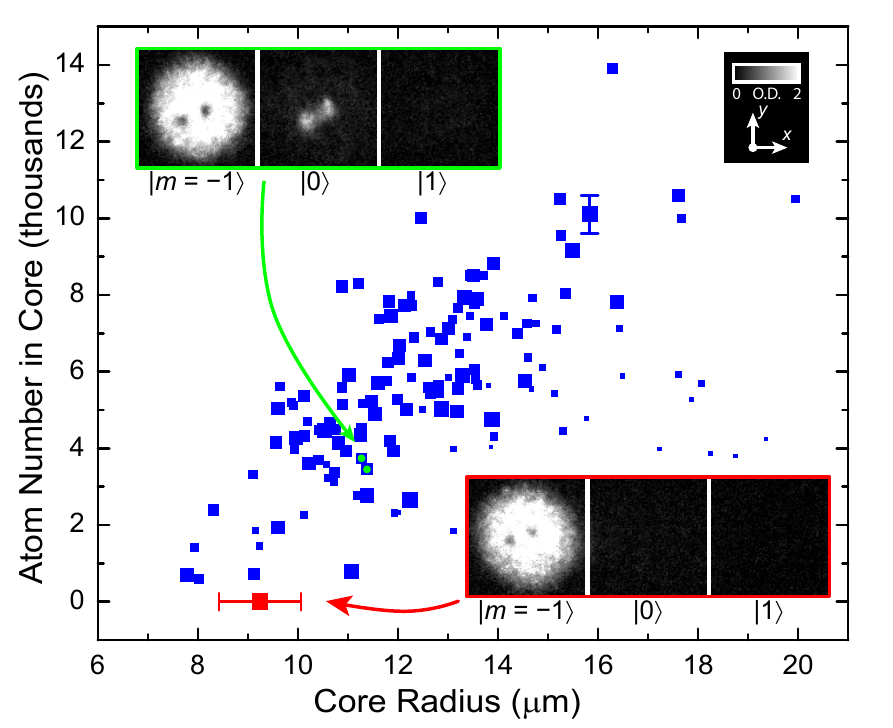}
\caption{\textbf{Effect of the core atoms on the size of the core.}
The post-expansion size of the vortex cores in the $m=-1$ component with unfilled (red) or filled (blue) cores. The core size depends in part on the number of $\ket{0}$ atoms within the core, which grows as $\ket{0}$ atoms accumulate there. Each blue point represents a single vortex measurement. As even empty vortex cores near the condensate boundary are enlarged, we indicate the radial position of a vortex by the size of the point, with smaller points corresponding to larger radii. A typical uncertainty in the atom number within the core is given as a vertical error bar for a single point. The red point and error bar illustrate the mean and standard deviation of a representative sample of unfilled vortex cores from more than 25~condensates, produced with a low-speed ($|\mathrm{d}B_\mathrm{b}/\mathrm{d}t| < 1~\mathrm{G~s}^{-1}$) ramp of the magnetic bias field. The insets show typical experimental atomic density profiles from absorption imaging, expressed in terms of dimensionless optical depth (O.D.), of the three spinor components for the case of filled (upper left, and points marked with green circles) and empty (lower right) cores. The field of view of each image in the insets is $219~\micron \times 219~\micron$. Source data are provided as a Source Data file.}
\label{fig:coresize}
\end{figure}

\textbf{Vortex core filling and interface.} For comparison, we also produce vortices with empty cores by reducing the ramp rate such that the spins rotate nearly adiabatically, leaving the system with unobservable populations in $\ket{0}$ and $\ket{1}$.  The size of the filled vortex core is typically much larger than that of an empty core, as shown in Fig.~\ref{fig:coresize}. We have numerically verified that for our experimental parameters  the superfluid vortex core expands at a rate similar to that of the whole condensate after the release from the trap. In the experiment, the size of the filled vortex core is a further manifestation of the topology of the spinor where the spinor interactions break the $\absF=1$ spin condition of the FM phase. In the ground state, the size of a filled vortex core is determined by a spin healing length\cite{ruostekoski_prl_2003,lovegrove_pra_2012} arising only from the spin--spin interactions, which  is much larger than the density healing length that limits the size of an empty core. Thus, as the condensate evolves, dissipation causes the filled vortex cores to inflate as $\ket{0}$ atoms accumulate there. We observe no corresponding growth of empty vortex cores, as also shown in Fig.~\ref{fig:coresize}.



Whereas the $\SO(3)$ order parameter of the FM phase may be represented by the orientation of an orthonormal vector triad, the \polar\ order parameter is characterised by an unoriented nematic axis $\nematic$ together with the condensate phase (Supplementary Note~2). The filling of the vortex core thus results in an interface between regions where the superfluid order parameter breaks different symmetries. In our system the interface appears in the internal structure of the defect itself and is observed directly in the experiment as a smooth transition between the FM vortex state in the surrounding superfluid and the \polar\ phase at the vortex core (Fig.~\ref{fig:mainresult}). A numerical simulation of this transition allows us to portray the condensate spinor graphically in terms of a spherical-harmonic expansion, $Z$ (see Fig.~\ref{fig:newintroduction}). The deformation of $Z$ illustrates the continuous topological interface that connects the $\SO(3)$ symmetric order parameter of the FM phase to the nematic order parameter of the \polar\ phase. Note that in the pure FM phase, the triad order parameter corresponds exactly to the orientation and argument of $Z$.

Analytically, the spinor describing the vortex and its superfluid core can be constructed as an interpolating filled-core vortex solution as in Ref.~\onlinecite{lovegrove_pra_2016},
\begin{equation}
  \label{eq:spinvortex-general}
  \zeta= \frac{1}{2}
     \threevec{\sqrt{2}e^{-i \phi}
                \left(\cos^2\frac{\beta}{2}D_+ - \sin^2\frac{\beta}{2}D_-\right)}
	      {\sin\beta\left(D_+ + D_-\right)}
	      {\sqrt{2}e^{i \phi}
		\left(\sin^2\frac{\beta}{2}D_+ - \cos^2\frac{\beta}{2}D_-\right)},
\end{equation}
where $D_\pm=\sqrt{1 \pm F}$ represents the interpolation between the FM and \polar\ phases for $F$ varying from 1 to~0, respectively.  The azimuthal angle around the vortex line is represented by $\phi$, and $\beta$ is the polar angle. The spin vector is $\expF = F(\sin\beta\rhohat + \cos\beta\zhat)$, and the unit vector orthogonal to it is $\nematic=-\cos\beta\rhohat+\sin\beta\zhat$, where $\rhohat$ is the radial unit vector relative to the vortex line. For $F=1$, equation~\eqref{eq:spinvortex-general} reduces to the singular FM vortex, and for $F=0$, the spinor represents the non-circulating \polar\ phase that occupies the vortex core.

\begin{figure*}
\centering
\includegraphics[width=0.8\linewidth]{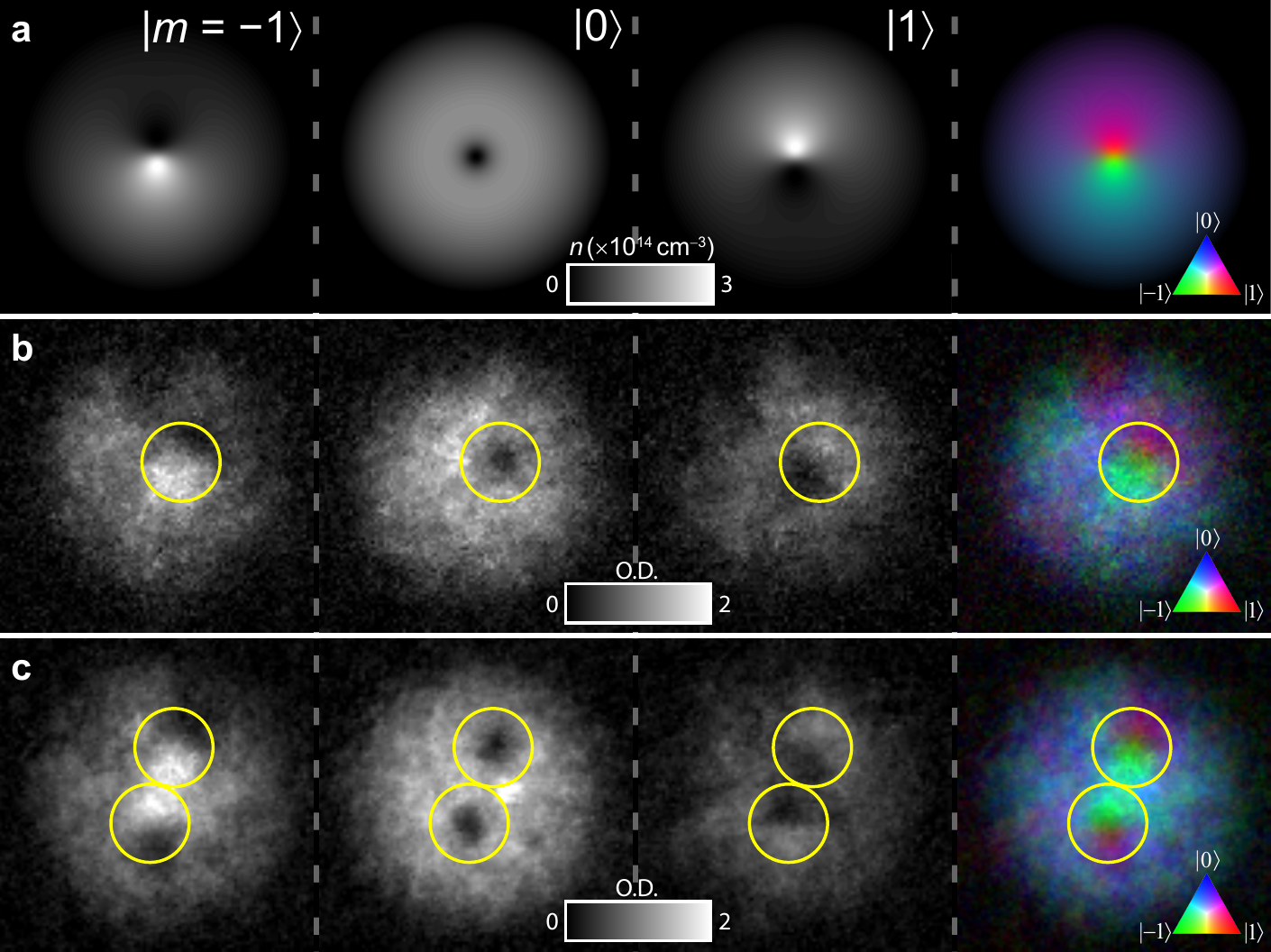}
\caption{\textbf{Signature of SO(3) character.}
(\textbf{a}) Spinor component densities after applying a $\pi/2$ spin-tip pulse to the analytically constructed singly quantized vortex, equation~\protect\eqref{eq:two-comp}, corresponding to a change of spinor basis. The vortex core is at the density minimum of the $\ket{0}$ component.
(\textbf{b,c}) Experimental absorption images of the atomic density, expressed in terms of dimensionless optical depth (O.D.), in each spinor component after applying a $\pi/2$ spin rotation for condensates containing one and two vortices, respectively. The vortex cores are identified by the density minima in the $m=0$ component, and yellow circles are drawn around the corresponding locations in each spinor component and in the colour composite image. The false-colour composite images show alternating regions of $m=\pm 1$ components in the vicinity of the vortex core. The field of view of each image in \textbf{b,c} is $219~\micron \times 219~\micron$. Source data are provided as a Source Data file.}
\label{fig:pi-2-pulse}
\end{figure*}

\textbf{Spinor analysis.}  Next, we explicitly demonstrate the $\SO(3)$ nature of the vortex. The representation of the vortex wave function as a three-component spinor depends on the choice of the spinor basis, and the order parameter symmetry dictates how the representation transforms under a change of basis. Experimentally it is more convenient to change the orientation of the spin with respect to a fixed quantization axis by applying a radio-frequency (RF) $\pi/2$ pulse, which rotates the spin according to the unitary transformation $U(0,\pi/2,\gamma_0)$ where the arbitrary angle $\gamma_0$ does not affect the outcome. The resulting density profiles are notably more complicated, as shown in Fig.~\ref{fig:pi-2-pulse}. To understand these results theoretically, we assume cylindrical symmetry and neglect any small population in $\ket{1}$, leading to a qualitative model for the vortex
\begin{equation}
  \label{eq:two-comp}
  \zeta =
    \threevec{0}
             {\sqrt{1-g(\rho)}}
             {e^{i\phi}\sqrt{g(\rho)}},
\end{equation}
where $g(\rho) = \rho^2/(\rho^2+r_0^2)$ approximates the vortex-core profile with size parameterised by $r_0$. The FM part of the spinor~\eqref{eq:two-comp} in the original basis transforms as $e^{i\phi}(0,0,1)^\mathrm{T} \to e^{i\phi}(1/2,-1/\sqrt{2},1/2)^\mathrm{T}$, distributing the atoms across all three components. The \polar\ part transforms as $(0,1,0)^\mathrm{T} \to (-1/\sqrt{2},0,1/\sqrt{2})^\mathrm{T}$, splitting the atoms evenly between the $\ket{\pm1}$ components. Thus, after the pulse, the original atomic density distribution of the FM phase is reproduced in the $\ket{0}$ component as it only contains atoms that originated in the $\ket{-1}$ component. On the other hand, the other two components exhibit phase singularities that have shifted to different locations, leading to a split-core solution that appears to have broken the axial symmetry of the original state. This translation of the vortices after the basis transformation is a manifestation of the $\SO(3)$ symmetry of the order parameter, and indicates the presence of a line singularity about which the spin vector rotates (disgyration). After the $\pi/2$ rotation, one can still identify the locations of the vortices by the density minima of the atoms in the $\ket{0}$ component.



The matter wave in $\ket{\pm1}$ may also be interpreted as an interference between the overlapping spinor components before the spin-tip pulse. In all cases, the experimental density profiles of Fig.~\ref{fig:pi-2-pulse} agree well with the theoretical prediction obtained by applying a $\pi/2$ spin rotation to equation~\eqref{eq:two-comp}.

\begin{center}
\textbf{\large Discussion}
\end{center}

Our results advance the experimental and theoretical investigations of defects containing topological interfaces. Similar techniques can be used to generate half-quantum vortices, as well as vortices with coherent interfaces involving the many diverse magnetic phases observed in spin-2 spinor condensates\cite{semenoff_prl_2007,kobayashi_prl_2009,mawson_pra_2015,borgh_prl_2016}. The filled  vortex cores themselves may be used as tracers to examine the longitudinal dynamics of the vortex lines\cite{fonda_pnas_2014}, which are otherwise difficult to discern. A further exciting extension would be to study the corresponding system in rotation where the nucleation and stability of vortices should dramatically depend on the precise value of the conserved magnetisation\cite{lovegrove_pra_2016} --- determining whether non-singular or singular vortices will prevail.


\begin{center}
\textbf{\large Methods}
\end{center}

\textbf{Experiment.} The experimental techniques resemble those described in Ref.~\citen{ray_nature_2014}, beginning with an optically trapped $^{87}$Rb condensate prepared in the FM phase $(1,0,0)^\mathrm{T} = \ket{1}$. The optical trap frequencies are $\omega_r \simeq 2\pi \times 130$~Hz and $\omega_z \simeq 2\pi \times 170$~Hz in the radial and axial directions, respectively, and with an initial atom number $N$ of typically $2\times 10^5$. The axial Thomas--Fermi radius of the condensate is 5~$\mu$m and the corresponding radial extent is 7~$\mu$m. The bias magnetic field $B_\mathrm{b}$ is controlled by a single Helmholtz coil pair, and the quadrupole magnetic field strength $b_\mathrm{q}$ by a second coaxial anti-Helmholtz pair. Two other pairs of coils for the $x$ and $y$ directions null those field components such that the field zero passes through the centre of the condensate.

The magnetic field zero is initially placed approximately 35~$\mu$m above the condensate with an initial gradient strength $b_\mathrm{q} =4.3(4)~\mathrm{G~cm}^{-1}$ and initial bias field $B_\mathrm{b} \approx 30$~mG. The bias field is then reduced to $\sim -50$~mG at the rate $\mathrm{d}B_\mathrm{b}/\mathrm{d}t$, and then to $-0.38$~G over the following 10~ms. The atoms are then held in the trap for a time $T_\mathrm{evolve}$. An optional $8~\mu$s, $0.266$~MHz RF $\pi/2$ spin-tip pulse is applied immediately afterwards. At the conclusion of the experiment the quadrupole field and the optical trap are extinguished. A brief exposure to a magnetic field gradient of $70~\mathrm{G~cm}^{-1}$ during the 23~ms expansion separates the spinor components horizontally, after which they are imaged absorptively along the $y$- and $z$-axes in a 0.1~G field aligned with the $z$-axis. Atom loss during the experiment, both during the ramp and during the subsequent evolution time, reduces the total number of atoms to approximately $2\times 10^5$ at the time of imaging.

Reducing the bias field at the rate $-0.25~\mathrm{G~s}^{-1}$ results in a doubly-quantized vortex in $\ket{-1}$ and essentially no atoms in the other spinor components. The experiments with filled cores were conducted at higher ramp rates, between $-4~\mathrm{G~s}^{-1}$ and $-6~\mathrm{G~s}^{-1}$. Ramp rates exceeding $-10~\mathrm{G~s}^{-1}$ result in larger non-singular vortices that occupy all three spinor components. These are not observed to evolve into singular $\SO(3)$ vortices.

\textbf{Numerical model.} We use experimental parameters for the Gross-Pitaevskii Hamiltonian density of
the spin-1 BEC
\begin{equation}
  \label{eq:hamiltonian-density}
\mathcal{H}
  =  h_0
    + \frac{c_0}{2}n^2
    + \frac{c_2}{2}n^2\absF^2
    - pn\inleva{\mathbf{B}\cdot\hat{\mathbf{F}}} + qn\inleva{(\mathbf{B}\cdot\hat{\mathbf{F}})^2},
\end{equation}
where $h_0 = \frac{\hbar^2}{2M_a}\abs{\nabla\Psi}^2 + V(\rr)n$ includes the harmonic trapping potential $V(\rr)$. Here $\hbar$ is the reduced Planck constant and $M_a$ is the atomic mass.
The spin  is defined as the expectation value $\expF=\sum_{\alpha\beta}\zeta^{\dagger}_\alpha\mathbf{\hat{F}}_{\alpha\beta}\zeta_\beta$, where $\mathbf{\hat{F}}$ is a vector of dimensionless spin-1 Pauli matrices.
The condensate spin vector corresponding to equation~\eqref{eq:general-fm} is given by
$\expF=\cos\alpha\sin\beta\xhat+\sin\alpha\sin\beta\yhat+\cos\beta\zhat$.
The FM order parameter can be defined by the orientation of two orthogonal vectors $\expF$ and
$\nematic =
(-\sin\alpha\cos\gamma-\cos\alpha\sin\gamma\cos\beta)\xhat+(\cos\alpha\cos\gamma-\sin\alpha\sin\gamma\cos\beta)\yhat+\sin\gamma\sin\beta\zhat$.
The last two terms of equation~\eqref{eq:hamiltonian-density} describe the linear and quadratic Zeeman shift of strengths $p$ and $q$, respectively. The two
interaction terms of strengths $c_0$ and $c_2$ arise from $s$-wave
scattering of the atoms.

In $s$-wave scattering the only spin-flip processes are $2\left|m=0\right> \rightleftharpoons \left|m=+1\right> + \left|m=-1\right>$. The longitudinal magnetisation
\begin{equation}
  M = \frac{1}{N} \int d^3r\,n(\rr)F_z(\rr),
\end{equation}
where $F_z$ is the $z$ component of the condensate spin,
is therefore approximately conserved on time scales for which $s$-wave scattering dominates. This condition is broken when the Gross--Pitaevskii equations are made dissipative, e.g., by imaginary-time evolution. We employ an algorithm to strictly restore the conservation of magnetization\cite{lovegrove_pra_2016}  throughout energy relaxation in pure imaginary time evolution and in evolution dynamics following imprinting, in which case we set time to include a small imaginary component $t \to (1-i\eta)t$, where $\eta \sim 10^{-2}$. All numerical simulations are carried out using a split-step algorithm on a minimum of $128\times128\times128$-point grid.

\begin{center}
\textbf{\large Data Availability}
\end{center}

All relevant datasets generated during and/or analysed during the current study are available from the corresponding author upon request.  The source data underlying Figs.~1 and~3--7 are provided as a Source Data file in the Zenodo repository (doi:19.5281/zenodo.3404017) (Ref.~\onlinecite{weiss_data_2019}).

\addvspace{\baselineskip}
\begin{center}
\textbf{\large References}
\end{center}
\addvspace{\baselineskip}

\begin{thebibliography}{46}%
\makeatletter
\providecommand \@ifxundefined [1]{%
 \@ifx{#1\undefined}
}%
\providecommand \@ifnum [1]{%
 \ifnum #1\expandafter \@firstoftwo
 \else \expandafter \@secondoftwo
 \fi
}%
\providecommand \@ifx [1]{%
 \ifx #1\expandafter \@firstoftwo
 \else \expandafter \@secondoftwo
 \fi
}%
\providecommand \natexlab [1]{#1}%
\providecommand \enquote  [1]{``#1''}%
\providecommand \bibnamefont  [1]{#1}%
\providecommand \bibfnamefont [1]{#1}%
\providecommand \citenamefont [1]{#1}%
\providecommand \href@noop [0]{\@secondoftwo}%
\providecommand \href [0]{\begingroup \@sanitize@url \@href}%
\providecommand \@href[1]{\@@startlink{#1}\@@href}%
\providecommand \@@href[1]{\endgroup#1\@@endlink}%
\providecommand \@sanitize@url [0]{\catcode `\\12\catcode `\$12\catcode
  `\&12\catcode `\#12\catcode `\^12\catcode `\_12\catcode `\%12\relax}%
\providecommand \@@startlink[1]{}%
\providecommand \@@endlink[0]{}%
\providecommand \url  [0]{\begingroup\@sanitize@url \@url }%
\providecommand \@url [1]{\endgroup\@href {#1}{\urlprefix }}%
\providecommand \urlprefix  [0]{URL }%
\providecommand \Eprint [0]{\href }%
\providecommand \doibase [0]{http://dx.doi.org/}%
\providecommand \selectlanguage [0]{\@gobble}%
\providecommand \bibinfo  [0]{\@secondoftwo}%
\providecommand \bibfield  [0]{\@secondoftwo}%
\providecommand \translation [1]{[#1]}%
\providecommand \BibitemOpen [0]{}%
\providecommand \bibitemStop [0]{}%
\providecommand \bibitemNoStop [0]{.\EOS\space}%
\providecommand \EOS [0]{\spacefactor3000\relax}%
\providecommand \BibitemShut  [1]{\csname bibitem#1\endcsname}%
\let\auto@bib@innerbib\@empty
\bibitem [{\citenamefont {Volovik}(2003)}]{volovik}%
  \BibitemOpen
  \bibfield  {author} {\bibinfo {author} {\bibfnamefont {Grigory~E.}\
  \bibnamefont {Volovik}},\ }\href@noop {} {\emph {\bibinfo {title} {The
  Universe in a Helium Droplet}}}\ (\bibinfo  {publisher} {Oxford University
  Press},\ \bibinfo {year} {2003})\BibitemShut {NoStop}%
\bibitem [{\citenamefont {Vollhardt}\ and\ \citenamefont
  {W\"olfle}(1990)}]{vollhardt-wolfle}%
  \BibitemOpen
  \bibfield  {author} {\bibinfo {author} {\bibfnamefont {Dieter}\ \bibnamefont
  {Vollhardt}}\ and\ \bibinfo {author} {\bibfnamefont {Peter}\ \bibnamefont
  {W\"olfle}},\ }\href@noop {} {\emph {\bibinfo {title} {The Superfluid Phases
  of Helium 3}}}\ (\bibinfo  {publisher} {Taylor {\&} Francis Ltd},\ \bibinfo
  {address} {London, UK},\ \bibinfo {year} {1990})\BibitemShut {NoStop}%
\bibitem [{\citenamefont {Kawaguchi}\ and\ \citenamefont
  {Ueda}(2012)}]{kawaguchi_physrep_2012}%
  \BibitemOpen
  \bibfield  {author} {\bibinfo {author} {\bibfnamefont {Yuki}\ \bibnamefont
  {Kawaguchi}}\ and\ \bibinfo {author} {\bibfnamefont {Masahito}\ \bibnamefont
  {Ueda}},\ }\bibfield  {title} {\enquote {\bibinfo {title} {Spinor
  {B}ose--{E}instein condensates},}\ }\href {\doibase
  doi:10.1016/j.physrep.2012.07.005} {\bibfield  {journal} {\bibinfo  {journal}
  {Phys. Rep.}\ }\textbf {\bibinfo {volume} {520}},\ \bibinfo {pages}
  {253--382} (\bibinfo {year} {2012})}\BibitemShut {NoStop}%
\bibitem [{\citenamefont {Stamper-Kurn}\ and\ \citenamefont
  {Ueda}(2013)}]{stamper-kurn_rmp_2013}%
  \BibitemOpen
  \bibfield  {author} {\bibinfo {author} {\bibfnamefont {Dan~M.}\ \bibnamefont
  {Stamper-Kurn}}\ and\ \bibinfo {author} {\bibfnamefont {Masahito}\
  \bibnamefont {Ueda}},\ }\bibfield  {title} {\enquote {\bibinfo {title}
  {Spinor {B}ose gases: Symmetries, magnetism, and quantum dynamics},}\ }\href
  {\doibase 10.1103/RevModPhys.85.1191} {\bibfield  {journal} {\bibinfo
  {journal} {Rev. Mod. Phys.}\ }\textbf {\bibinfo {volume} {85}},\ \bibinfo
  {pages} {1191--1244} (\bibinfo {year} {2013})}\BibitemShut {NoStop}%
\bibitem [{\citenamefont {Donnelly}(1991)}]{donnelly}%
  \BibitemOpen
  \bibfield  {author} {\bibinfo {author} {\bibfnamefont {Russell~J.}\
  \bibnamefont {Donnelly}},\ }\href@noop {} {\emph {\bibinfo {title} {Quantized
  Vortices in Helium {II}}}}\ (\bibinfo  {publisher} {Cambridge University
  Press, Cambridge},\ \bibinfo {year} {1991})\BibitemShut {NoStop}%
\bibitem [{\citenamefont {Fetter}(2009)}]{fetter_rmp_2009}%
  \BibitemOpen
  \bibfield  {author} {\bibinfo {author} {\bibfnamefont {Alexander~L.}\
  \bibnamefont {Fetter}},\ }\bibfield  {title} {\enquote {\bibinfo {title}
  {Rotating trapped {B}ose--{E}instein condensates},}\ }\href {\doibase
  10.1103/RevModPhys.81.647} {\bibfield  {journal} {\bibinfo  {journal} {Rev.
  Mod. Phys.}\ }\textbf {\bibinfo {volume} {81}},\ \bibinfo {pages} {647--691}
  (\bibinfo {year} {2009})}\BibitemShut {NoStop}%
\bibitem [{\citenamefont {Mermin}(1979)}]{mermin_rmp_1979}%
  \BibitemOpen
  \bibfield  {author} {\bibinfo {author} {\bibfnamefont {N.~D.}\ \bibnamefont
  {Mermin}},\ }\bibfield  {title} {\enquote {\bibinfo {title} {The topological
  theory of defects in ordered media},}\ }\href {\doibase
  10.1103/RevModPhys.51.591} {\bibfield  {journal} {\bibinfo  {journal} {Rev.
  Mod. Phys.}\ }\textbf {\bibinfo {volume} {51}},\ \bibinfo {pages} {591--648}
  (\bibinfo {year} {1979})}\BibitemShut {NoStop}%
\bibitem [{\citenamefont {Staley}(2010)}]{staley_ejp_2010}%
  \BibitemOpen
  \bibfield  {author} {\bibinfo {author} {\bibfnamefont {Mark}\ \bibnamefont
  {Staley}},\ }\bibfield  {title} {\enquote {\bibinfo {title} {Understanding
  quaternions and the {D}irac belt trick},}\ }\href {\doibase
  doi:10.1088/0143-0807/31/3/004} {\bibfield  {journal} {\bibinfo  {journal}
  {Eur. J. Phys.}\ }\textbf {\bibinfo {volume} {31}},\ \bibinfo {pages}
  {467--478} (\bibinfo {year} {2010})}\BibitemShut {NoStop}%
\bibitem [{\citenamefont {Simola}\ \emph {et~al.}(1987)\citenamefont {Simola},
  \citenamefont {Skrbek}, \citenamefont {Nummila},\ and\ \citenamefont
  {Korhonen}}]{Simola87}%
  \BibitemOpen
  \bibfield  {author} {\bibinfo {author} {\bibfnamefont {J.~T.}\ \bibnamefont
  {Simola}}, \bibinfo {author} {\bibfnamefont {L.}~\bibnamefont {Skrbek}},
  \bibinfo {author} {\bibfnamefont {K.~K.}\ \bibnamefont {Nummila}}, \ and\
  \bibinfo {author} {\bibfnamefont {J.~S.}\ \bibnamefont {Korhonen}},\
  }\bibfield  {title} {\enquote {\bibinfo {title} {Two different vortex states
  in rotating $^{3}${H}e--\textit{A} observed by use of negative ions},}\
  }\href {\doibase 10.1103/PhysRevLett.58.904} {\bibfield  {journal} {\bibinfo
  {journal} {Phys. Rev. Lett.}\ }\textbf {\bibinfo {volume} {58}},\ \bibinfo
  {pages} {904--907} (\bibinfo {year} {1987})}\BibitemShut {NoStop}%
\bibitem [{\citenamefont {Parts}\ \emph {et~al.}(1995)\citenamefont {Parts},
  \citenamefont {Karim\"aki}, \citenamefont {Koivuniemi}, \citenamefont
  {Krusius}, \citenamefont {Ruutu}, \citenamefont {Thuneberg},\ and\
  \citenamefont {Volovik}}]{parts_prl_1995}%
  \BibitemOpen
  \bibfield  {author} {\bibinfo {author} {\bibfnamefont {{\"U}.}~\bibnamefont
  {Parts}}, \bibinfo {author} {\bibfnamefont {J.~M.}\ \bibnamefont
  {Karim\"aki}}, \bibinfo {author} {\bibfnamefont {J.~H.}\ \bibnamefont
  {Koivuniemi}}, \bibinfo {author} {\bibfnamefont {M.}~\bibnamefont {Krusius}},
  \bibinfo {author} {\bibfnamefont {V.~M.~H.}\ \bibnamefont {Ruutu}}, \bibinfo
  {author} {\bibfnamefont {E.~V.}\ \bibnamefont {Thuneberg}}, \ and\ \bibinfo
  {author} {\bibfnamefont {G.~E.}\ \bibnamefont {Volovik}},\ }\bibfield
  {title} {\enquote {\bibinfo {title} {Phase diagram of vortices in superfluid
  ${}^{3}${H}e--$\mathit{A}$},}\ }\href {\doibase 10.1103/PhysRevLett.75.3320}
  {\bibfield  {journal} {\bibinfo  {journal} {Phys. Rev. Lett.}\ }\textbf
  {\bibinfo {volume} {75}},\ \bibinfo {pages} {3320--3323} (\bibinfo {year}
  {1995})}\BibitemShut {NoStop}%
\bibitem [{\citenamefont {Isoshima}\ and\ \citenamefont
  {Machida}(2002)}]{isoshima_pra_2002}%
  \BibitemOpen
  \bibfield  {author} {\bibinfo {author} {\bibfnamefont {Tomoya}\ \bibnamefont
  {Isoshima}}\ and\ \bibinfo {author} {\bibfnamefont {Kazushige}\ \bibnamefont
  {Machida}},\ }\bibfield  {title} {\enquote {\bibinfo {title} {Axisymmetric
  vortices in spinor {B}ose--{E}instein condensates under rotation},}\ }\href
  {\doibase 10.1103/PhysRevA.66.023602} {\bibfield  {journal} {\bibinfo
  {journal} {Phys. Rev. A}\ }\textbf {\bibinfo {volume} {66}},\ \bibinfo
  {pages} {023602} (\bibinfo {year} {2002})}\BibitemShut {NoStop}%
\bibitem [{\citenamefont {Lovegrove}\ \emph {et~al.}(2012)\citenamefont
  {Lovegrove}, \citenamefont {Borgh},\ and\ \citenamefont
  {Ruostekoski}}]{lovegrove_pra_2012}%
  \BibitemOpen
  \bibfield  {author} {\bibinfo {author} {\bibfnamefont {Justin}\ \bibnamefont
  {Lovegrove}}, \bibinfo {author} {\bibfnamefont {Magnus~O.}\ \bibnamefont
  {Borgh}}, \ and\ \bibinfo {author} {\bibfnamefont {Janne}\ \bibnamefont
  {Ruostekoski}},\ }\bibfield  {title} {\enquote {\bibinfo {title}
  {Energetically stable singular vortex cores in an atomic spin-1
  {B}ose--{E}instein condensate},}\ }\href {\doibase
  10.1103/PhysRevA.86.013613} {\bibfield  {journal} {\bibinfo  {journal} {Phys.
  Rev. A}\ }\textbf {\bibinfo {volume} {86}},\ \bibinfo {pages} {013613}
  (\bibinfo {year} {2012})}\BibitemShut {NoStop}%
\bibitem [{\citenamefont {Kobayashi}\ \emph {et~al.}(2012)\citenamefont
  {Kobayashi}, \citenamefont {Kawaguchi}, \citenamefont {Nitta},\ and\
  \citenamefont {Ueda}}]{kobayashi_pra_2012}%
  \BibitemOpen
  \bibfield  {author} {\bibinfo {author} {\bibfnamefont {Shingo}\ \bibnamefont
  {Kobayashi}}, \bibinfo {author} {\bibfnamefont {Yuki}\ \bibnamefont
  {Kawaguchi}}, \bibinfo {author} {\bibfnamefont {Muneto}\ \bibnamefont
  {Nitta}}, \ and\ \bibinfo {author} {\bibfnamefont {Masahito}\ \bibnamefont
  {Ueda}},\ }\bibfield  {title} {\enquote {\bibinfo {title} {Topological
  classification of vortex-core structures of spin-1 {B}ose--{E}instein
  condensates},}\ }\href {\doibase 10.1103/PhysRevA.86.023612} {\bibfield
  {journal} {\bibinfo  {journal} {Phys. Rev. A}\ }\textbf {\bibinfo {volume}
  {86}},\ \bibinfo {pages} {023612} (\bibinfo {year} {2012})}\BibitemShut
  {NoStop}%
\bibitem [{\citenamefont {Lovegrove}\ \emph {et~al.}(2016)\citenamefont
  {Lovegrove}, \citenamefont {Borgh},\ and\ \citenamefont
  {Ruostekoski}}]{lovegrove_pra_2016}%
  \BibitemOpen
  \bibfield  {author} {\bibinfo {author} {\bibfnamefont {Justin}\ \bibnamefont
  {Lovegrove}}, \bibinfo {author} {\bibfnamefont {Magnus~O.}\ \bibnamefont
  {Borgh}}, \ and\ \bibinfo {author} {\bibfnamefont {Janne}\ \bibnamefont
  {Ruostekoski}},\ }\bibfield  {title} {\enquote {\bibinfo {title} {Stability
  and internal structure of vortices in spin-1 {B}ose--{E}instein condensates
  with conserved magnetization},}\ }\href {\doibase 10.1103/PhysRevA.93.033633}
  {\bibfield  {journal} {\bibinfo  {journal} {Phys. Rev. A}\ }\textbf {\bibinfo
  {volume} {93}},\ \bibinfo {pages} {033633} (\bibinfo {year}
  {2016})}\BibitemShut {NoStop}%
\bibitem [{\citenamefont {Ruostekoski}\ and\ \citenamefont
  {Anglin}(2003)}]{ruostekoski_prl_2003}%
  \BibitemOpen
  \bibfield  {author} {\bibinfo {author} {\bibfnamefont {J.}~\bibnamefont
  {Ruostekoski}}\ and\ \bibinfo {author} {\bibfnamefont {J.~R.}\ \bibnamefont
  {Anglin}},\ }\bibfield  {title} {\enquote {\bibinfo {title} {Monopole core
  instability and {A}lice rings in spinor {B}ose--{E}instein condensates},}\
  }\href {\doibase 10.1103/PhysRevLett.91.190402} {\bibfield  {journal}
  {\bibinfo  {journal} {Phys. Rev. Lett.}\ }\textbf {\bibinfo {volume} {91}},\
  \bibinfo {pages} {190402} (\bibinfo {year} {2003})}\BibitemShut {NoStop}%
\bibitem [{\citenamefont {Sadler}\ \emph {et~al.}(2006)\citenamefont {Sadler},
  \citenamefont {Higbie}, \citenamefont {Leslie}, \citenamefont
  {Vengalattore},\ and\ \citenamefont {Stamper-Kurn}}]{sadler_nature_2006}%
  \BibitemOpen
  \bibfield  {author} {\bibinfo {author} {\bibfnamefont {L.~E.}\ \bibnamefont
  {Sadler}}, \bibinfo {author} {\bibfnamefont {J.~M.}\ \bibnamefont {Higbie}},
  \bibinfo {author} {\bibfnamefont {S.~R.}\ \bibnamefont {Leslie}}, \bibinfo
  {author} {\bibfnamefont {M.}~\bibnamefont {Vengalattore}}, \ and\ \bibinfo
  {author} {\bibfnamefont {D.~M.}\ \bibnamefont {Stamper-Kurn}},\ }\bibfield
  {title} {\enquote {\bibinfo {title} {Spontaneous symmetry breaking in a
  quenched ferromagnetic spinor {B}ose--{E}instein condensate},}\ }\href
  {\doibase doi:10.1038/nature05094} {\bibfield  {journal} {\bibinfo  {journal}
  {Nature}\ }\textbf {\bibinfo {volume} {443}},\ \bibinfo {pages} {312--315}
  (\bibinfo {year} {2006})}\BibitemShut {NoStop}%
\bibitem [{\citenamefont {Chen}\ \emph {et~al.}(2018)\citenamefont {Chen},
  \citenamefont {Liu}, \citenamefont {Tsai}, \citenamefont {Chiu},
  \citenamefont {Kawaguchi}, \citenamefont {Yip}, \citenamefont {Chang},\ and\
  \citenamefont {Lin}}]{chen_prl_2018}%
  \BibitemOpen
  \bibfield  {author} {\bibinfo {author} {\bibfnamefont {P.-K.}\ \bibnamefont
  {Chen}}, \bibinfo {author} {\bibfnamefont {L.-R.}\ \bibnamefont {Liu}},
  \bibinfo {author} {\bibfnamefont {M.-J.}\ \bibnamefont {Tsai}}, \bibinfo
  {author} {\bibfnamefont {N.-C.}\ \bibnamefont {Chiu}}, \bibinfo {author}
  {\bibfnamefont {Y.}~\bibnamefont {Kawaguchi}}, \bibinfo {author}
  {\bibfnamefont {S.-K.}\ \bibnamefont {Yip}}, \bibinfo {author} {\bibfnamefont
  {M.-S.}\ \bibnamefont {Chang}}, \ and\ \bibinfo {author} {\bibfnamefont
  {Y.-J.}\ \bibnamefont {Lin}},\ }\bibfield  {title} {\enquote {\bibinfo
  {title} {Rotating atomic quantum gases with light-induced azimuthal gauge
  potentials and the observation of the {H}ess--{F}airbank effect},}\ }\href
  {\doibase 10.1103/PhysRevLett.121.250401} {\bibfield  {journal} {\bibinfo
  {journal} {Phys. Rev. Lett.}\ }\textbf {\bibinfo {volume} {121}},\ \bibinfo
  {pages} {250401} (\bibinfo {year} {2018})}\BibitemShut {NoStop}%
\bibitem [{\citenamefont {Seo}\ \emph {et~al.}(2015)\citenamefont {Seo},
  \citenamefont {Kang}, \citenamefont {Kwon},\ and\ \citenamefont
  {Shin}}]{seo_prl_2015}%
  \BibitemOpen
  \bibfield  {author} {\bibinfo {author} {\bibfnamefont {Sang~Won}\
  \bibnamefont {Seo}}, \bibinfo {author} {\bibfnamefont {Seji}\ \bibnamefont
  {Kang}}, \bibinfo {author} {\bibfnamefont {Woo~Jin}\ \bibnamefont {Kwon}}, \
  and\ \bibinfo {author} {\bibfnamefont {Yong-il}\ \bibnamefont {Shin}},\
  }\bibfield  {title} {\enquote {\bibinfo {title} {Half-quantum vortices in an
  antiferromagnetic spinor {B}ose--{E}instein condensate},}\ }\href {\doibase
  10.1103/PhysRevLett.115.015301} {\bibfield  {journal} {\bibinfo  {journal}
  {Phys. Rev. Lett.}\ }\textbf {\bibinfo {volume} {115}},\ \bibinfo {pages}
  {015301} (\bibinfo {year} {2015})}\BibitemShut {NoStop}%
\bibitem [{\citenamefont {Borgh}\ and\ \citenamefont
  {Ruostekoski}(2012)}]{borgh_prl_2012}%
  \BibitemOpen
  \bibfield  {author} {\bibinfo {author} {\bibfnamefont {Magnus~O.}\
  \bibnamefont {Borgh}}\ and\ \bibinfo {author} {\bibfnamefont {Janne}\
  \bibnamefont {Ruostekoski}},\ }\bibfield  {title} {\enquote {\bibinfo {title}
  {Topological interface engineering and defect crossing in ultracold atomic
  gases},}\ }\href {\doibase 10.1103/PhysRevLett.109.015302} {\bibfield
  {journal} {\bibinfo  {journal} {Phys. Rev. Lett.}\ }\textbf {\bibinfo
  {volume} {109}},\ \bibinfo {pages} {015302} (\bibinfo {year}
  {2012})}\BibitemShut {NoStop}%
\bibitem [{\citenamefont {Finne}\ \emph {et~al.}(2006)\citenamefont {Finne},
  \citenamefont {Eltsov}, \citenamefont {H\"{a}nninen}, \citenamefont {Kopnin},
  \citenamefont {Kopu}, \citenamefont {Krusius}, \citenamefont {Tsubota},\ and\
  \citenamefont {Volovik}}]{finne_rpp_2006}%
  \BibitemOpen
  \bibfield  {author} {\bibinfo {author} {\bibfnamefont {A.~P.}\ \bibnamefont
  {Finne}}, \bibinfo {author} {\bibfnamefont {V.~B.}\ \bibnamefont {Eltsov}},
  \bibinfo {author} {\bibfnamefont {R.}~\bibnamefont {H\"{a}nninen}}, \bibinfo
  {author} {\bibfnamefont {N.~B.}\ \bibnamefont {Kopnin}}, \bibinfo {author}
  {\bibfnamefont {J.}~\bibnamefont {Kopu}}, \bibinfo {author} {\bibfnamefont
  {M.}~\bibnamefont {Krusius}}, \bibinfo {author} {\bibfnamefont
  {M.}~\bibnamefont {Tsubota}}, \ and\ \bibinfo {author} {\bibfnamefont
  {G.~E.}\ \bibnamefont {Volovik}},\ }\bibfield  {title} {\enquote {\bibinfo
  {title} {Dynamics of vortices and interfaces in superfluid {$^3$He}},}\
  }\href@noop {} {\bibfield  {journal} {\bibinfo  {journal} {Rep. Prog. Phys.}\
  }\textbf {\bibinfo {volume} {69}},\ \bibinfo {pages} {3157--3230} (\bibinfo
  {year} {2006})}\BibitemShut {NoStop}%
\bibitem [{\citenamefont {Bradley}\ \emph {et~al.}(2008)\citenamefont
  {Bradley}, \citenamefont {Fisher}, \citenamefont {Guenault}, \citenamefont
  {Haley}, \citenamefont {Kopu}, \citenamefont {Martin}, \citenamefont
  {Pickett}, \citenamefont {Roberts},\ and\ \citenamefont
  {Tsepelin}}]{bradley_nphys_2008}%
  \BibitemOpen
  \bibfield  {author} {\bibinfo {author} {\bibfnamefont {D.~I.}\ \bibnamefont
  {Bradley}}, \bibinfo {author} {\bibfnamefont {S.~N.}\ \bibnamefont {Fisher}},
  \bibinfo {author} {\bibfnamefont {A.~M.}\ \bibnamefont {Guenault}}, \bibinfo
  {author} {\bibfnamefont {R.~P.}\ \bibnamefont {Haley}}, \bibinfo {author}
  {\bibfnamefont {J.}~\bibnamefont {Kopu}}, \bibinfo {author} {\bibfnamefont
  {H.}~\bibnamefont {Martin}}, \bibinfo {author} {\bibfnamefont {G.~R.}\
  \bibnamefont {Pickett}}, \bibinfo {author} {\bibfnamefont {J.~E.}\
  \bibnamefont {Roberts}}, \ and\ \bibinfo {author} {\bibfnamefont
  {V.}~\bibnamefont {Tsepelin}},\ }\bibfield  {title} {\enquote {\bibinfo
  {title} {Relic topological defects from brane annihilation simulated in
  superfluid $^3${H}e},}\ }\href {https://www.nature.com/articles/nphys815}
  {\bibfield  {journal} {\bibinfo  {journal} {Nat. Phys.}\ }\textbf {\bibinfo
  {volume} {4}},\ \bibinfo {pages} {46--49} (\bibinfo {year}
  {2008})}\BibitemShut {NoStop}%
\bibitem [{\citenamefont {Kibble}(1976)}]{kibble_jpa_1976}%
  \BibitemOpen
  \bibfield  {author} {\bibinfo {author} {\bibfnamefont {T.~W.~B.}\
  \bibnamefont {Kibble}},\ }\bibfield  {title} {\enquote {\bibinfo {title}
  {Topology of cosmic domains and strings},}\ }\href {\doibase
  10.1088/0305-4470/9/8/029} {\bibfield  {journal} {\bibinfo  {journal} {J.
  Phys. A Math. Gen.}\ }\textbf {\bibinfo {volume} {9}},\ \bibinfo {pages}
  {1387--1398} (\bibinfo {year} {1976})}\BibitemShut {NoStop}%
\bibitem [{\citenamefont {Sarangi}\ and\ \citenamefont
  {Tye}(2002)}]{sarangi_plb_2002}%
  \BibitemOpen
  \bibfield  {author} {\bibinfo {author} {\bibfnamefont {Saswat}\ \bibnamefont
  {Sarangi}}\ and\ \bibinfo {author} {\bibfnamefont {S.-H.~Henry}\ \bibnamefont
  {Tye}},\ }\bibfield  {title} {\enquote {\bibinfo {title} {Cosmic string
  production towards the end of brane inflation},}\ }\href
  {https://www.sciencedirect.com/science/article/pii/S0370269302018245}
  {\bibfield  {journal} {\bibinfo  {journal} {Phys. Lett. B}\ }\textbf
  {\bibinfo {volume} {536}},\ \bibinfo {pages} {185--192} (\bibinfo {year}
  {2002})}\BibitemShut {NoStop}%
\bibitem [{\citenamefont {Bert}\ \emph {et~al.}(2011)\citenamefont {Bert},
  \citenamefont {Kalisky}, \citenamefont {Bell}, \citenamefont {Kim},
  \citenamefont {Hikita}, \citenamefont {Hwang},\ and\ \citenamefont
  {Moler}}]{bert_nphys_2011}%
  \BibitemOpen
  \bibfield  {author} {\bibinfo {author} {\bibfnamefont {Julie~A.}\
  \bibnamefont {Bert}}, \bibinfo {author} {\bibfnamefont {Beena}\ \bibnamefont
  {Kalisky}}, \bibinfo {author} {\bibfnamefont {Christopher}\ \bibnamefont
  {Bell}}, \bibinfo {author} {\bibfnamefont {Minu}\ \bibnamefont {Kim}},
  \bibinfo {author} {\bibfnamefont {Yasuyuki}\ \bibnamefont {Hikita}}, \bibinfo
  {author} {\bibfnamefont {Harold~Y.}\ \bibnamefont {Hwang}}, \ and\ \bibinfo
  {author} {\bibfnamefont {Kathryn~A.}\ \bibnamefont {Moler}},\ }\bibfield
  {title} {\enquote {\bibinfo {title} {Direct imaging of the coexistence of
  ferromagnetism and superconductivity at the {LaAlO3/SrTiO3} interface},}\
  }\href {https://www.nature.com/articles/nphys2079} {\bibfield  {journal}
  {\bibinfo  {journal} {Nat. Phys.}\ }\textbf {\bibinfo {volume} {7}},\
  \bibinfo {pages} {767--771} (\bibinfo {year} {2011})}\BibitemShut {NoStop}%
\bibitem [{\citenamefont {Leanhardt}\ \emph {et~al.}(2003)\citenamefont
  {Leanhardt}, \citenamefont {Shin}, \citenamefont {Kielpinski}, \citenamefont
  {Pritchard},\ and\ \citenamefont {Ketterle}}]{leanhardt_prl_2003}%
  \BibitemOpen
  \bibfield  {author} {\bibinfo {author} {\bibfnamefont {A.~E.}\ \bibnamefont
  {Leanhardt}}, \bibinfo {author} {\bibfnamefont {Y.}~\bibnamefont {Shin}},
  \bibinfo {author} {\bibfnamefont {D.}~\bibnamefont {Kielpinski}}, \bibinfo
  {author} {\bibfnamefont {D.~E.}\ \bibnamefont {Pritchard}}, \ and\ \bibinfo
  {author} {\bibfnamefont {W.}~\bibnamefont {Ketterle}},\ }\bibfield  {title}
  {\enquote {\bibinfo {title} {Coreless vortex formation in a spinor
  {B}ose--{E}instein condensate},}\ }\href {\doibase
  10.1103/PhysRevLett.90.140403} {\bibfield  {journal} {\bibinfo  {journal}
  {Phys. Rev. Lett.}\ }\textbf {\bibinfo {volume} {90}},\ \bibinfo {pages}
  {140403} (\bibinfo {year} {2003})}\BibitemShut {NoStop}%
\bibitem [{\citenamefont {Leslie}\ \emph {et~al.}(2009)\citenamefont {Leslie},
  \citenamefont {Hansen}, \citenamefont {Wright}, \citenamefont {Deutsch},\
  and\ \citenamefont {Bigelow}}]{leslie_prl_2009}%
  \BibitemOpen
  \bibfield  {author} {\bibinfo {author} {\bibfnamefont {L.~S.}\ \bibnamefont
  {Leslie}}, \bibinfo {author} {\bibfnamefont {A.}~\bibnamefont {Hansen}},
  \bibinfo {author} {\bibfnamefont {K.~C.}\ \bibnamefont {Wright}}, \bibinfo
  {author} {\bibfnamefont {B.~M.}\ \bibnamefont {Deutsch}}, \ and\ \bibinfo
  {author} {\bibfnamefont {N.~P.}\ \bibnamefont {Bigelow}},\ }\bibfield
  {title} {\enquote {\bibinfo {title} {Creation and detection of skyrmions in a
  {B}ose--{E}instein condensate},}\ }\href {\doibase
  10.1103/PhysRevLett.103.250401} {\bibfield  {journal} {\bibinfo  {journal}
  {Phys. Rev. Lett.}\ }\textbf {\bibinfo {volume} {103}},\ \bibinfo {pages}
  {250401} (\bibinfo {year} {2009})}\BibitemShut {NoStop}%
\bibitem [{\citenamefont {Choi}\ \emph
  {et~al.}(2012{\natexlab{a}})\citenamefont {Choi}, \citenamefont {Kwon},\ and\
  \citenamefont {Shin}}]{choi_prl_2012}%
  \BibitemOpen
  \bibfield  {author} {\bibinfo {author} {\bibfnamefont {Jae-yoon}\
  \bibnamefont {Choi}}, \bibinfo {author} {\bibfnamefont {Woo~Jin}\
  \bibnamefont {Kwon}}, \ and\ \bibinfo {author} {\bibfnamefont {Yong-il}\
  \bibnamefont {Shin}},\ }\bibfield  {title} {\enquote {\bibinfo {title}
  {Observation of topologically stable 2{D} skyrmions in an antiferromagnetic
  spinor {B}ose--{E}instein condensate},}\ }\href {\doibase
  10.1103/PhysRevLett.108.035301} {\bibfield  {journal} {\bibinfo  {journal}
  {Phys. Rev. Lett.}\ }\textbf {\bibinfo {volume} {108}},\ \bibinfo {pages}
  {035301} (\bibinfo {year} {2012}{\natexlab{a}})}\BibitemShut {NoStop}%
\bibitem [{\citenamefont {Leanhardt}\ \emph {et~al.}(2002)\citenamefont
  {Leanhardt}, \citenamefont {G\"orlitz}, \citenamefont {Chikkatur},
  \citenamefont {Kielpinski}, \citenamefont {Shin}, \citenamefont {Pritchard},\
  and\ \citenamefont {Ketterle}}]{leanhardt_prl_2002}%
  \BibitemOpen
  \bibfield  {author} {\bibinfo {author} {\bibfnamefont {A.~E.}\ \bibnamefont
  {Leanhardt}}, \bibinfo {author} {\bibfnamefont {A.}~\bibnamefont
  {G\"orlitz}}, \bibinfo {author} {\bibfnamefont {A.~P.}\ \bibnamefont
  {Chikkatur}}, \bibinfo {author} {\bibfnamefont {D.}~\bibnamefont
  {Kielpinski}}, \bibinfo {author} {\bibfnamefont {Y.}~\bibnamefont {Shin}},
  \bibinfo {author} {\bibfnamefont {D.~E.}\ \bibnamefont {Pritchard}}, \ and\
  \bibinfo {author} {\bibfnamefont {W.}~\bibnamefont {Ketterle}},\ }\bibfield
  {title} {\enquote {\bibinfo {title} {Imprinting vortices in a
  {B}ose--{E}instein condensate using topological phases},}\ }\href@noop {}
  {\bibfield  {journal} {\bibinfo  {journal} {Phys. Rev. Lett.}\ }\textbf
  {\bibinfo {volume} {89}},\ \bibinfo {pages} {190403} (\bibinfo {year}
  {2002})}\BibitemShut {NoStop}%
\bibitem [{\citenamefont {Shin}\ \emph {et~al.}(2004)\citenamefont {Shin},
  \citenamefont {Saba}, \citenamefont {Vengalattore}, \citenamefont {Pasquini},
  \citenamefont {Sanner}, \citenamefont {Leanhardt}, \citenamefont {Prentiss},
  \citenamefont {Pritchard},\ and\ \citenamefont {Ketterle}}]{shin_prl_2004}%
  \BibitemOpen
  \bibfield  {author} {\bibinfo {author} {\bibfnamefont {Y.}~\bibnamefont
  {Shin}}, \bibinfo {author} {\bibfnamefont {M.}~\bibnamefont {Saba}}, \bibinfo
  {author} {\bibfnamefont {M.}~\bibnamefont {Vengalattore}}, \bibinfo {author}
  {\bibfnamefont {T.~A.}\ \bibnamefont {Pasquini}}, \bibinfo {author}
  {\bibfnamefont {C.}~\bibnamefont {Sanner}}, \bibinfo {author} {\bibfnamefont
  {A.~E.}\ \bibnamefont {Leanhardt}}, \bibinfo {author} {\bibfnamefont
  {M.}~\bibnamefont {Prentiss}}, \bibinfo {author} {\bibfnamefont {D.~E.}\
  \bibnamefont {Pritchard}}, \ and\ \bibinfo {author} {\bibfnamefont
  {W.}~\bibnamefont {Ketterle}},\ }\bibfield  {title} {\enquote {\bibinfo
  {title} {Dynamical instability of a doubly quantized vortex in a
  {B}ose--{E}instein condensate},}\ }\href {\doibase
  10.1103/PhysRevLett.93.160406} {\bibfield  {journal} {\bibinfo  {journal}
  {Phys. Rev. Lett.}\ }\textbf {\bibinfo {volume} {93}},\ \bibinfo {pages}
  {160406} (\bibinfo {year} {2004})}\BibitemShut {NoStop}%
\bibitem [{\citenamefont {Huhtam\"aki}\ \emph {et~al.}(2006)\citenamefont
  {Huhtam\"aki}, \citenamefont {M\"ott\"onen}, \citenamefont {Isoshima},
  \citenamefont {Pietil\"a},\ and\ \citenamefont
  {Virtanen}}]{huhtamaki_prl_2006}%
  \BibitemOpen
  \bibfield  {author} {\bibinfo {author} {\bibfnamefont {J.~A.~M.}\
  \bibnamefont {Huhtam\"aki}}, \bibinfo {author} {\bibfnamefont
  {M.}~\bibnamefont {M\"ott\"onen}}, \bibinfo {author} {\bibfnamefont
  {T.}~\bibnamefont {Isoshima}}, \bibinfo {author} {\bibfnamefont
  {V.}~\bibnamefont {Pietil\"a}}, \ and\ \bibinfo {author} {\bibfnamefont
  {S.~M.~M.}\ \bibnamefont {Virtanen}},\ }\bibfield  {title} {\enquote
  {\bibinfo {title} {Splitting times of doubly quantized vortices in dilute
  {B}ose--{E}instein condensates},}\ }\href {\doibase
  10.1103/PhysRevLett.97.110406} {\bibfield  {journal} {\bibinfo  {journal}
  {Phys. Rev. Lett.}\ }\textbf {\bibinfo {volume} {97}},\ \bibinfo {pages}
  {110406} (\bibinfo {year} {2006})}\BibitemShut {NoStop}%
\bibitem [{\citenamefont {Isoshima}\ \emph {et~al.}(2007)\citenamefont
  {Isoshima}, \citenamefont {Okano}, \citenamefont {Yasuda}, \citenamefont
  {Kasa}, \citenamefont {Huhtam\"aki}, \citenamefont {Kumakura},\ and\
  \citenamefont {Takahashi}}]{isoshima_prl_2007}%
  \BibitemOpen
  \bibfield  {author} {\bibinfo {author} {\bibfnamefont {T.}~\bibnamefont
  {Isoshima}}, \bibinfo {author} {\bibfnamefont {M.}~\bibnamefont {Okano}},
  \bibinfo {author} {\bibfnamefont {H.}~\bibnamefont {Yasuda}}, \bibinfo
  {author} {\bibfnamefont {K.}~\bibnamefont {Kasa}}, \bibinfo {author}
  {\bibfnamefont {J.~A.~M.}\ \bibnamefont {Huhtam\"aki}}, \bibinfo {author}
  {\bibfnamefont {M.}~\bibnamefont {Kumakura}}, \ and\ \bibinfo {author}
  {\bibfnamefont {Y.}~\bibnamefont {Takahashi}},\ }\bibfield  {title} {\enquote
  {\bibinfo {title} {Spontaneous splitting of a quadruply charged vortex},}\
  }\href {\doibase 10.1103/PhysRevLett.99.200403} {\bibfield  {journal}
  {\bibinfo  {journal} {Phys. Rev. Lett.}\ }\textbf {\bibinfo {volume} {99}},\
  \bibinfo {pages} {200403} (\bibinfo {year} {2007})}\BibitemShut {NoStop}%
\bibitem [{\citenamefont {Nakahara}\ \emph {et~al.}(2000)\citenamefont
  {Nakahara}, \citenamefont {Isoshima}, \citenamefont {Machida}, \citenamefont
  {Ogawa},\ and\ \citenamefont {Ohmi}}]{NAKAHARA00}%
  \BibitemOpen
  \bibfield  {author} {\bibinfo {author} {\bibfnamefont {Mikio}\ \bibnamefont
  {Nakahara}}, \bibinfo {author} {\bibfnamefont {Tomoya}\ \bibnamefont
  {Isoshima}}, \bibinfo {author} {\bibfnamefont {Kazushige}\ \bibnamefont
  {Machida}}, \bibinfo {author} {\bibfnamefont {Shin-ichiro}\ \bibnamefont
  {Ogawa}}, \ and\ \bibinfo {author} {\bibfnamefont {Tetsuo}\ \bibnamefont
  {Ohmi}},\ }\bibfield  {title} {\enquote {\bibinfo {title} {A simple method to
  create a vortex in {B}ose--{E}instein condensate of alkali atoms},}\ }\href
  {\doibase https://doi.org/10.1016/S0921-4526(99)01952-3} {\bibfield
  {journal} {\bibinfo  {journal} {Physica B}\ }\textbf {\bibinfo {volume}
  {284--288}},\ \bibinfo {pages} {17--18} (\bibinfo {year} {2000})}\BibitemShut
  {NoStop}%
\bibitem [{\citenamefont {Pietil\"a}\ and\ \citenamefont
  {M\"ott\"onen}(2009)}]{pietila_prl_2009_dirac}%
  \BibitemOpen
  \bibfield  {author} {\bibinfo {author} {\bibfnamefont {Ville}\ \bibnamefont
  {Pietil\"a}}\ and\ \bibinfo {author} {\bibfnamefont {Mikko}\ \bibnamefont
  {M\"ott\"onen}},\ }\bibfield  {title} {\enquote {\bibinfo {title} {Creation
  of {D}irac monopoles in spinor {B}ose--{E}instein condensates},}\ }\href
  {\doibase 10.1103/PhysRevLett.103.030401} {\bibfield  {journal} {\bibinfo
  {journal} {Phys. Rev. Lett.}\ }\textbf {\bibinfo {volume} {103}},\ \bibinfo
  {pages} {030401} (\bibinfo {year} {2009})}\BibitemShut {NoStop}%
\bibitem [{\citenamefont {Choi}\ \emph
  {et~al.}(2012{\natexlab{b}})\citenamefont {Choi}, \citenamefont {Kwon},
  \citenamefont {Lee}, \citenamefont {Jeong}, \citenamefont {An},\ and\
  \citenamefont {Shin}}]{choi_njp_2012}%
  \BibitemOpen
  \bibfield  {author} {\bibinfo {author} {\bibfnamefont {Jae-yoon}\
  \bibnamefont {Choi}}, \bibinfo {author} {\bibfnamefont {Woo~Jin}\
  \bibnamefont {Kwon}}, \bibinfo {author} {\bibfnamefont {Moonjoo}\
  \bibnamefont {Lee}}, \bibinfo {author} {\bibfnamefont {Hyunseok}\
  \bibnamefont {Jeong}}, \bibinfo {author} {\bibfnamefont {Kyungwon}\
  \bibnamefont {An}}, \ and\ \bibinfo {author} {\bibfnamefont {Yong-il}\
  \bibnamefont {Shin}},\ }\bibfield  {title} {\enquote {\bibinfo {title}
  {Imprinting skyrmion spin textures in {B}ose--{E}instein condensates},}\
  }\href {http://stacks.iop.org/1367-2630/14/i=5/a=053013} {\bibfield
  {journal} {\bibinfo  {journal} {New J. Phys.}\ }\textbf {\bibinfo {volume}
  {14}},\ \bibinfo {pages} {053013} (\bibinfo {year}
  {2012}{\natexlab{b}})}\BibitemShut {NoStop}%
\bibitem [{\citenamefont {Ray}\ \emph {et~al.}(2014)\citenamefont {Ray},
  \citenamefont {Ruokokoski}, \citenamefont {Kandel}, \citenamefont
  {M\"ott\"onen},\ and\ \citenamefont {Hall}}]{ray_nature_2014}%
  \BibitemOpen
  \bibfield  {author} {\bibinfo {author} {\bibfnamefont {M.~W.}\ \bibnamefont
  {Ray}}, \bibinfo {author} {\bibfnamefont {E.}~\bibnamefont {Ruokokoski}},
  \bibinfo {author} {\bibfnamefont {S.}~\bibnamefont {Kandel}}, \bibinfo
  {author} {\bibfnamefont {M.}~\bibnamefont {M\"ott\"onen}}, \ and\ \bibinfo
  {author} {\bibfnamefont {D.~S.}\ \bibnamefont {Hall}},\ }\bibfield  {title}
  {\enquote {\bibinfo {title} {Observation of {D}irac monopoles in a synthetic
  magnetic field},}\ }\href {\doibase 10.1038/nature12954} {\bibfield
  {journal} {\bibinfo  {journal} {Nature}\ }\textbf {\bibinfo {volume} {505}},\
  \bibinfo {pages} {657--660} (\bibinfo {year} {2014})}\BibitemShut {NoStop}%
\bibitem [{\citenamefont {Lee}\ \emph {et~al.}(2018)\citenamefont {Lee},
  \citenamefont {Gheorghe}, \citenamefont {Tiurev}, \citenamefont {Ollikainen},
  \citenamefont {M\"ott\"onen},\ and\ \citenamefont {Hall}}]{lee_sciadv_2018}%
  \BibitemOpen
  \bibfield  {author} {\bibinfo {author} {\bibfnamefont {W.}~\bibnamefont
  {Lee}}, \bibinfo {author} {\bibfnamefont {A.~H.}\ \bibnamefont {Gheorghe}},
  \bibinfo {author} {\bibfnamefont {K.}~\bibnamefont {Tiurev}}, \bibinfo
  {author} {\bibfnamefont {T.}~\bibnamefont {Ollikainen}}, \bibinfo {author}
  {\bibfnamefont {M.}~\bibnamefont {M\"ott\"onen}}, \ and\ \bibinfo {author}
  {\bibfnamefont {D.~S.}\ \bibnamefont {Hall}},\ }\bibfield  {title} {\enquote
  {\bibinfo {title} {Synthetic electromagnetic knot in a three-dimensional
  skyrmion},}\ }\href {\doibase 10.1126/sciadv.aao3820} {\bibfield  {journal}
  {\bibinfo  {journal} {Sci. Adv.}\ }\textbf {\bibinfo {volume} {4}},\ \bibinfo
  {pages} {eaao3820} (\bibinfo {year} {2018})}\BibitemShut {NoStop}%
\bibitem [{\citenamefont {Hall}\ \emph {et~al.}(2016)\citenamefont {Hall},
  \citenamefont {Ray}, \citenamefont {Tiurev}, \citenamefont {Ruokokoski},
  \citenamefont {Gheorghe},\ and\ \citenamefont
  {M{\"o}tt{\"o}nen}}]{hall_nphys_2016}%
  \BibitemOpen
  \bibfield  {author} {\bibinfo {author} {\bibfnamefont {D.~S.}\ \bibnamefont
  {Hall}}, \bibinfo {author} {\bibfnamefont {M.~W.}\ \bibnamefont {Ray}},
  \bibinfo {author} {\bibfnamefont {K.}~\bibnamefont {Tiurev}}, \bibinfo
  {author} {\bibfnamefont {E.}~\bibnamefont {Ruokokoski}}, \bibinfo {author}
  {\bibfnamefont {A.~H.}\ \bibnamefont {Gheorghe}}, \ and\ \bibinfo {author}
  {\bibfnamefont {M.}~\bibnamefont {M{\"o}tt{\"o}nen}},\ }\bibfield  {title}
  {\enquote {\bibinfo {title} {Tying quantum knots},}\ }\href {\doibase
  10.1038/nphys3624} {\bibfield  {journal} {\bibinfo  {journal} {Nat. Phys.}\
  }\textbf {\bibinfo {volume} {12}},\ \bibinfo {pages} {478--483} (\bibinfo
  {year} {2016})}\BibitemShut {NoStop}%
\bibitem [{\citenamefont {Matthews}\ \emph {et~al.}(1999)\citenamefont
  {Matthews}, \citenamefont {Anderson}, \citenamefont {Haljan}, \citenamefont
  {Hall}, \citenamefont {Wieman},\ and\ \citenamefont
  {Cornell}}]{matthews_prl_1999}%
  \BibitemOpen
  \bibfield  {author} {\bibinfo {author} {\bibfnamefont {M.~R.}\ \bibnamefont
  {Matthews}}, \bibinfo {author} {\bibfnamefont {B.~P.}\ \bibnamefont
  {Anderson}}, \bibinfo {author} {\bibfnamefont {P.~C.}\ \bibnamefont
  {Haljan}}, \bibinfo {author} {\bibfnamefont {D.~S.}\ \bibnamefont {Hall}},
  \bibinfo {author} {\bibfnamefont {C.~E.}\ \bibnamefont {Wieman}}, \ and\
  \bibinfo {author} {\bibfnamefont {E.~A.}\ \bibnamefont {Cornell}},\
  }\bibfield  {title} {\enquote {\bibinfo {title} {Vortices in a
  {B}ose--{E}instein condensate},}\ }\href {\doibase
  10.1103/PhysRevLett.83.2498} {\bibfield  {journal} {\bibinfo  {journal}
  {Phys. Rev. Lett.}\ }\textbf {\bibinfo {volume} {83}},\ \bibinfo {pages}
  {2498--2501} (\bibinfo {year} {1999})}\BibitemShut {NoStop}%
\bibitem [{\citenamefont {Andersen}\ \emph {et~al.}(2006)\citenamefont
  {Andersen}, \citenamefont {Ryu}, \citenamefont {Clad\'e}, \citenamefont
  {Natarajan}, \citenamefont {Vaziri}, \citenamefont {Helmerson},\ and\
  \citenamefont {Phillips}}]{andersen_prl_2006}%
  \BibitemOpen
  \bibfield  {author} {\bibinfo {author} {\bibfnamefont {M.~F.}\ \bibnamefont
  {Andersen}}, \bibinfo {author} {\bibfnamefont {C.}~\bibnamefont {Ryu}},
  \bibinfo {author} {\bibfnamefont {Pierre}\ \bibnamefont {Clad\'e}}, \bibinfo
  {author} {\bibfnamefont {Vasant}\ \bibnamefont {Natarajan}}, \bibinfo
  {author} {\bibfnamefont {A.}~\bibnamefont {Vaziri}}, \bibinfo {author}
  {\bibfnamefont {K.}~\bibnamefont {Helmerson}}, \ and\ \bibinfo {author}
  {\bibfnamefont {W.~D.}\ \bibnamefont {Phillips}},\ }\bibfield  {title}
  {\enquote {\bibinfo {title} {Quantized rotation of atoms from photons with
  orbital angular momentum},}\ }\href {\doibase 10.1103/PhysRevLett.97.170406}
  {\bibfield  {journal} {\bibinfo  {journal} {Phys. Rev. Lett.}\ }\textbf
  {\bibinfo {volume} {97}},\ \bibinfo {pages} {170406} (\bibinfo {year}
  {2006})}\BibitemShut {NoStop}%
\bibitem [{\citenamefont {Rosenbusch}\ \emph {et~al.}(2002)\citenamefont
  {Rosenbusch}, \citenamefont {Bretin},\ and\ \citenamefont
  {Dalibard}}]{rosenbusch_prl_2002}%
  \BibitemOpen
  \bibfield  {author} {\bibinfo {author} {\bibfnamefont {P.}~\bibnamefont
  {Rosenbusch}}, \bibinfo {author} {\bibfnamefont {V.}~\bibnamefont {Bretin}},
  \ and\ \bibinfo {author} {\bibfnamefont {J.}~\bibnamefont {Dalibard}},\
  }\bibfield  {title} {\enquote {\bibinfo {title} {Dynamics of a single vortex
  line in a {B}ose--{E}instein condensate},}\ }\href {\doibase
  10.1103/PhysRevLett.89.200403} {\bibfield  {journal} {\bibinfo  {journal}
  {Phys. Rev. Lett.}\ }\textbf {\bibinfo {volume} {89}},\ \bibinfo {pages}
  {200403} (\bibinfo {year} {2002})}\BibitemShut {NoStop}%
\bibitem [{\citenamefont {Semenoff}\ and\ \citenamefont
  {Zhou}(2007)}]{semenoff_prl_2007}%
  \BibitemOpen
  \bibfield  {author} {\bibinfo {author} {\bibfnamefont {Gordon~W.}\
  \bibnamefont {Semenoff}}\ and\ \bibinfo {author} {\bibfnamefont {Fei}\
  \bibnamefont {Zhou}},\ }\bibfield  {title} {\enquote {\bibinfo {title}
  {Discrete symmetries and $1/3$--quantum vortices in condensates of ${F}=2$
  cold atoms},}\ }\href {\doibase 10.1103/PhysRevLett.98.100401} {\bibfield
  {journal} {\bibinfo  {journal} {Phys. Rev. Lett.}\ }\textbf {\bibinfo
  {volume} {98}},\ \bibinfo {pages} {100401} (\bibinfo {year}
  {2007})}\BibitemShut {NoStop}%
\bibitem [{\citenamefont {Kobayashi}\ \emph {et~al.}(2009)\citenamefont
  {Kobayashi}, \citenamefont {Kawaguchi}, \citenamefont {Nitta},\ and\
  \citenamefont {Ueda}}]{kobayashi_prl_2009}%
  \BibitemOpen
  \bibfield  {author} {\bibinfo {author} {\bibfnamefont {Michikazu}\
  \bibnamefont {Kobayashi}}, \bibinfo {author} {\bibfnamefont {Yuki}\
  \bibnamefont {Kawaguchi}}, \bibinfo {author} {\bibfnamefont {Muneto}\
  \bibnamefont {Nitta}}, \ and\ \bibinfo {author} {\bibfnamefont {Masahito}\
  \bibnamefont {Ueda}},\ }\bibfield  {title} {\enquote {\bibinfo {title}
  {Collision dynamics and rung formation of non-{A}belian vortices},}\ }\href
  {\doibase 10.1103/PhysRevLett.103.115301} {\bibfield  {journal} {\bibinfo
  {journal} {Phys. Rev. Lett.}\ }\textbf {\bibinfo {volume} {103}},\ \bibinfo
  {pages} {115301} (\bibinfo {year} {2009})}\BibitemShut {NoStop}%
\bibitem [{\citenamefont {Mawson}\ \emph {et~al.}(2015)\citenamefont {Mawson},
  \citenamefont {Ruben},\ and\ \citenamefont {Simula}}]{mawson_pra_2015}%
  \BibitemOpen
  \bibfield  {author} {\bibinfo {author} {\bibfnamefont {Thomas}\ \bibnamefont
  {Mawson}}, \bibinfo {author} {\bibfnamefont {Gary}\ \bibnamefont {Ruben}}, \
  and\ \bibinfo {author} {\bibfnamefont {Tapio}\ \bibnamefont {Simula}},\
  }\bibfield  {title} {\enquote {\bibinfo {title} {Route to non-{A}belian
  quantum turbulence in spinor {B}ose--{E}instein condensates},}\ }\href
  {\doibase 10.1103/PhysRevA.91.063630} {\bibfield  {journal} {\bibinfo
  {journal} {Phys. Rev. A}\ }\textbf {\bibinfo {volume} {91}},\ \bibinfo
  {pages} {063630} (\bibinfo {year} {2015})}\BibitemShut {NoStop}%
\bibitem [{\citenamefont {Borgh}\ and\ \citenamefont
  {Ruostekoski}(2016)}]{borgh_prl_2016}%
  \BibitemOpen
  \bibfield  {author} {\bibinfo {author} {\bibfnamefont {Magnus~O.}\
  \bibnamefont {Borgh}}\ and\ \bibinfo {author} {\bibfnamefont {Janne}\
  \bibnamefont {Ruostekoski}},\ }\bibfield  {title} {\enquote {\bibinfo {title}
  {Core structure and non-{A}belian reconnection of defects in a biaxial
  nematic spin-2 {B}ose--{E}instein condensate},}\ }\href {\doibase
  10.1103/PhysRevLett.117.275302} {\bibfield  {journal} {\bibinfo  {journal}
  {Phys. Rev. Lett.}\ }\textbf {\bibinfo {volume} {117}},\ \bibinfo {pages}
  {275302} (\bibinfo {year} {2016})}\BibitemShut {NoStop}%
\bibitem [{\citenamefont {Fonda}\ \emph {et~al.}(2014)\citenamefont {Fonda},
  \citenamefont {Meichle}, \citenamefont {Ouellette}, \citenamefont {Hormoz},\
  and\ \citenamefont {Lathrop}}]{fonda_pnas_2014}%
  \BibitemOpen
  \bibfield  {author} {\bibinfo {author} {\bibfnamefont {Enrico}\ \bibnamefont
  {Fonda}}, \bibinfo {author} {\bibfnamefont {David~P.}\ \bibnamefont
  {Meichle}}, \bibinfo {author} {\bibfnamefont {Nicholas~T.}\ \bibnamefont
  {Ouellette}}, \bibinfo {author} {\bibfnamefont {Sahand}\ \bibnamefont
  {Hormoz}}, \ and\ \bibinfo {author} {\bibfnamefont {Daniel~P.}\ \bibnamefont
  {Lathrop}},\ }\bibfield  {title} {\enquote {\bibinfo {title} {Direct
  observation of {K}elvin waves excited by quantized vortex reconnection},}\
  }\href {\doibase 10.1073/pnas.1312536110} {\bibfield  {journal} {\bibinfo
  {journal} {Proc. Natl Acad. Sci.}\ }\textbf {\bibinfo {volume} {111}},\
  \bibinfo {pages} {4707--4710} (\bibinfo {year} {2014})}\BibitemShut {NoStop}%
\bibitem [{\citenamefont {Weiss}\ \emph {et~al.}(2019)\citenamefont {Weiss},
  \citenamefont {Borgh}, \citenamefont {Blinova}, \citenamefont {Ollikainen},
  \citenamefont {M\"ott\"onen}, \citenamefont {Ruostekoski},\ and\
  \citenamefont {Hall}}]{weiss_data_2019}%
  \BibitemOpen
  \bibfield  {author} {\bibinfo {author} {\bibfnamefont {L.~S.}\ \bibnamefont
  {Weiss}}, \bibinfo {author} {\bibfnamefont {M.~O.}\ \bibnamefont {Borgh}},
  \bibinfo {author} {\bibfnamefont {A.}~\bibnamefont {Blinova}}, \bibinfo
  {author} {\bibfnamefont {T.}~\bibnamefont {Ollikainen}}, \bibinfo {author}
  {\bibfnamefont {M.}~\bibnamefont {M\"ott\"onen}}, \bibinfo {author}
  {\bibfnamefont {J.}~\bibnamefont {Ruostekoski}}, \ and\ \bibinfo {author}
  {\bibfnamefont {D.S.}\ \bibnamefont {Hall}},\ }\href {\doibase
  10.5281/zenodo.3404017} {\enquote {\bibinfo {title} {Controlled creation of a
  singular spinor vortex by circumventing the {D}irac belt trick},}\ }\bibinfo
  {howpublished} {Zenodo digital repository. doi:10.5281/zenodo.340417}
  (\bibinfo {year} {2019})\BibitemShut {NoStop}%
\end{thebibliography}

\begin{thebibliography}{1}
\expandafter\ifx\csname url\endcsname\relax
  \def\url#1{\texttt{#1}}\fi
\expandafter\ifx\csname urlprefix\endcsname\relax\def\urlprefix{URL }\fi
\providecommand{\bibinfo}[2]{#2}
\providecommand{\eprint}[2][]{\url{#2}}

\bibitem{ho_prl_1998_s}
\bibinfo{author}{Ho, T.-L.}
\newblock \bibinfo{title}{Spinor {B}ose condensates in optical traps}.
\newblock \emph{\bibinfo{journal}{Phys. Rev. Lett.}}
  \textbf{\bibinfo{volume}{81}}, \bibinfo{pages}{742--745}
  (\bibinfo{year}{1998}).

\bibitem{lovegrove_pra_2016_s}
\bibinfo{author}{Lovegrove, J.}, \bibinfo{author}{Borgh, M.~O.} \&
  \bibinfo{author}{Ruostekoski, J.}
\newblock \bibinfo{title}{Stability and internal structure of vortices in
  spin-1 {B}ose--{E}instein condensates with conserved magnetization}.
\newblock \emph{\bibinfo{journal}{Phys. Rev. A}} \textbf{\bibinfo{volume}{93}},
  \bibinfo{pages}{033633} (\bibinfo{year}{2016}).

\bibitem{kawaguchi_physrep_2012_s}
\bibinfo{author}{Kawaguchi, Y.} \& \bibinfo{author}{Ueda, M.}
\newblock \bibinfo{title}{Spinor {B}ose--{E}instein condensates}.
\newblock \emph{\bibinfo{journal}{Phys. Rep.}}
  \textbf{\bibinfo{volume}{520}}, \bibinfo{pages}{253--382}
  (\bibinfo{year}{2012}).

\bibitem{seo_prl_2015_s}
\bibinfo{author}{Seo, S.~W.}, \bibinfo{author}{Kang, S.},
  \bibinfo{author}{Kwon, W.~J.} \& \bibinfo{author}{Shin, Y.-i.}
\newblock \bibinfo{title}{Half-quantum vortices in an antiferromagnetic spinor
  {B}ose--{E}instein condensate}.
\newblock \emph{\bibinfo{journal}{Phys. Rev. Lett.}}
  \textbf{\bibinfo{volume}{115}}, \bibinfo{pages}{015301}
  (\bibinfo{year}{2015}).

\end{thebibliography}


%

\addvspace{2\baselineskip}

\begin{center}
\textbf{\large Acknowledgements}
\end{center}

We acknowledge funding by the National Science Foundation (Grant Nos.\ PHY--1519174 and PHY--1806318), by the Emil Aaltonen Foundation, by the Kaupallisten ja teknillisten tieteiden tukis\"a\"ati\"o (KAUTE) foundation through its Researchers Abroad program, by the Academy of Finland through the Centre of Excellence Programme (grant no.\ 312300) and ERC Consolidator Grant QUESS (no.\ 681311), and by the EPSRC.

\begin{center}
\textbf{\large Author contributions}
\end{center}
L.S.W., A.B., T.O., and D.S.H.\ developed and conducted the experiments and analysed the data. J.R.\  proposed the experiment. Numerical simulations were carried out by M.O.B., and the theoretical analysis was developed by M.O.B.\ and J.R.
The manuscript was written by M.O.B., J.R.\ and D.S.H. All authors discussed the results and commented on the manuscript.


\begin{center}
\textbf{\large Additional information}
\end{center}

\textbf{\small Competing Interests:} The Authors declare no competing interests.

\textbf{Correspondence and requests for materials} should be addressed to D.S.H.\\ (\href{mailto:dshall@amherst.edu}{\texttt{dshall@amherst.edu}}).

\clearpage \widetext


\thispagestyle{empty}
\begin{center}
{\large\textbf{Controlled Creation of a Singular Spinor Vortex \\ by Circumventing the Dirac Belt Trick}}\par
Weiss et al. \par
\end{center}

\setcounter{page}{1}

\setcounter{figure}{0}
\setcounter{equation}{0}


\setcounter{page}{1}

\begin{center}
\textbf{\large Supplementary Notes}
\end{center}

\addvspace{\baselineskip}\noindent\textbf{Supplementary Note 1 $\vert$ Theoretical details of the $\SO(3)$ vortex creation process.} A non-singular vortex with longitudinal magnetisation $M$ can be written as
\begin{equation}
  \label{eq:cl-M}
  \zeta = \frac{1}{2}
    \threevec{1+Mf(\rho)}
             {\sqrt{2}e^{i\phi}\sqrt{1-[Mf(\rho)]^2}}
             {e^{2i\phi}[1-Mf(\rho)]},
\end{equation}
where $\phi$ is the azimuthal angle around the vortex line.
The condensate spin is $\expF = \sqrt{1-[Mf(\rho)]^2}\rhohat +
Mf(\rho)\zhat$ and the profile function $f(\rho)=\cos(\beta)/M$ satisfies \linebreak $\frac{1}{N}\int
d^3r\, n(\rr) f(\rho) = 1$. The analytically constructed spinor wave function for the non-singular vortex used in Figs.~\ref{fig:theory+data-densities}a and~\ref{fig:theory+data}a was obtained from supplementary equation~\eqref{eq:cl-M}
and embedded into the ground-state density profile of the condensate.

In these expressions $\beta(\rho)$ is the second Euler angle appearing in
equation~\eqref{eq:general-fm}, which in this case
increases monotonically from zero on the vortex line to $\pi$ at the
boundary to define a fountain-like spin texture.
For strongly negative $M$, corresponding to $\beta(\rho)$ increasing rapidly, a large proportion of the condensate resides in the $\ket{-1}$ component.
The two extremes of magnetisation correspond to constant
$f(\rho)$: For $M=1$, $f(\rho) = 1$, to give the vortex-free state
$\zeta=(1,0,0)^\mathrm{T}$. For $M=-1$, on the other hand, we
obtain the doubly-quantized
vortex $\zeta=(0,0,e^{2i\phi})^\mathrm{T}$ with its associated instability
towards splitting. The two limits can be continuously
connected through a family of functions $f(\rho)$, allowing the vortex
to unwind\citeS{ho_prl_1998_s}.
Note, however, that conservation of longitudinal magnetisation suppresses the redistribution of the atom populations between the spinor components during the free evolution of the condensate.

Using numerical energy relaxation (Methods), we find that if the vortex is prepared such that the condensate spin vector bends rapidly towards $-\zhat$ with increasing radial distance from the vortex centre, the spin interactions can no longer maintain a non-singular
profile while still preserving the spinor nature of the condensate.
The threshold approximately corresponds to a strong longitudinal
magnetisation $M\lesssim-0.3$ that is explicitly conserved, and the vortex decays by
splitting into a pair of singly quantized vortices as shown in
Figs.~\ref{fig:theory+data-densities} and~\ref{fig:theory+data}. We show the state reached
shortly after splitting of the initial vortex, exhibiting singular
vortices with fully formed, filled cores. This state is locally
stable: The singular vortices persist for a significant period of
imaginary-time evolution, but do eventually leave the
condensate over a significantly longer period of relaxation. The local
stability is also reflected in the long lifetime of the vortices
created in our experiment.

The observed splitting of the coreless vortex represents a particular
case of the instability also found in supplementary
Ref.~\onlinecite{lovegrove_pra_2016_s}. In imprinting of a non-singular
vortex, its magnetisation is a controllable parameter determined by
the rate of change of the imprinting magnetic field.
Our idealised model thus indicates
that singular $\SO(3)$ vortices can be created by imprinting a
non-singular vortex with sufficiently strong longitudinal
magnetisation and allowing it to split.

\noindent\textbf{Supplementary Note 2 $\vert$ Magnetic Phases.} The spin-1 Bose-Einstein condensate exhibits both polar (\polar) and ferromagnetic (FM) phases\citeS{kawaguchi_physrep_2012_s}. In a $^{87}$Rb condensate, $c_2<0$ in equation~\eqref{eq:hamiltonian-density}, and hence the FM phase, which maximises $\absF$, forms the ground state in a uniform system at zero magnetic field. The \polar\ phase, with $\absF=0$, appears in the cores of the singular vortices and exhibits different symmetry properties. The \polar\ order parameter, and hence also the spinor  $\zeta^{\rm\polar}$ in the \polar\ phase, are specified by a nematic axis $\nematic$ together with a condensate scalar phase $\tau$, with the property that $\zeta^{\rm\polar}(\nematic,\tau) = \zeta^{\rm\polar}(-\nematic, \tau + \pi)$. The \polar\ phase thus exhibits uniaxial nematic order, characterised by the continuous symmetry under spin rotations about the local, unoriented axis defined by $\nematic$. The order parameter space is therefore $[S^2\times\U(1)]/\mathbb{Z}_2$, where the factorisation by $\mathbb{Z}_2$ arises from the aforementioned nematic symmetry, which also allows the \polar\ phase to support half-quantum vortices\citeS{seo_prl_2015_s}.

\begin{center}
\textbf{\large Supplementary References}
\end{center}




\end{document}